\pdfoutput=1
\documentclass[printer]{rQUF2e}
\usepackage{subfigure}

\usepackage{url}
\usepackage{booktabs}

\theoremstyle{plain}
\newtheorem{theorem}{Theorem}[section]

\newtheorem{lemma}[theorem]{Lemma}
\newtheorem{proposition}[theorem]{Proposition}

\newtheorem{definition}[theorem]{Definition}

\theoremstyle{remark}

\usepackage[hidelinks]{hyperref}
\usepackage{cleveref}
\renewcommand{\ref}[1]{\labelcref{#1}}
\renewcommand{\eqref}[1]{\labelcref{#1}}
\hypersetup{
  pdftitle={ABIDES-MARL: A Multi-Agent Reinforcement Learning Environment for Optimal Execution with Endogenous Liquidity},
  pdfauthor={Patrick Cheridito, Jean-Loup Dupret, and Zhexin Wu},
  pdfsubject={Optimal execution under endogenous liquidity},
  pdfkeywords={optimal execution, endogenous liquidity, multi-agent reinforcement learning, market microstructure, limit order book}
}

\begin{document}

\title{ABIDES-MARL: A Multi-Agent Reinforcement Learning Environment for Optimal Execution with Endogenous Liquidity}

\author{Patrick Cheridito$\dag$, Jean-Loup Dupret$\dag$ and Zhexin Wu$^{\ast}$${\ddag}$\thanks{$^\ast$Corresponding author: \texttt{zhexin.wu@proton.me}}\\
\affil{$\dag$Department of Mathematics, ETH Zurich, Switzerland\\
$\ddag$Department of Management, Technology, and Economics, ETH Zurich, Switzerland} \received{} }

\maketitle
\thispagestyle{empty}

\begin{abstract}
Classical optimal execution models treat market impact as a pre-specified, exogenous process. However, when market makers adapt strategically, this assumption becomes a structural misspecification: execution dynamics depend on the policies and actions of other agents. The problem therefore ceases to be a single-agent control problem and instead becomes a finite-horizon stochastic game, in which liquidity emerges endogenously from the interactions among heterogeneous market players. We hence introduce \textbf{ABIDES-MARL}, a multi-agent reinforcement learning framework for studying optimal execution under endogenous liquidity in a realistic limit order book setting. The framework extends ABIDES-Gym to support multiple learning agents with synchronized decision periods that preserve proper information filtration and key market microstructure features. We validate the framework in an extended Kyle setting, where multiple learning agents recover gradual price discovery. Numerical results then show that execution strategies learned under endogenous, adaptive liquidity differ fundamentally from those implied by fixed exogenous price-impact benchmarks. In particular, strategies optimized under exogenous impact perform poorly once market makers adapt strategically: when information is not balanced across traders and market makers, market makers exploit predictable execution and the market dynamics may degenerate. The framework therefore provides a reproducible foundation for studying strategic adaptation in realistic markets and contributes to the development of economically interpretable agentic AI systems in finance.
\end{abstract}


\section{Introduction}

Optimal execution studies the problem of a liquidity trader who seeks to trade a large position over a finite horizon while minimizing transaction costs and controlling risk. In its classical formulation, the trader chooses a trading schedule that balances price impact against inventory risk. Early discrete-time models by \cite{BERTSIMAS1998} and the mean–variance framework of \cite{almgren_optimal_2001} formalized this trade-off and derived explicit optimal schedules under linear impact assumptions. Subsequent work incorporated time-varying liquidity \citep{Huberman2005Liquidity}, continuous-time stochastic control formulations \citep{cartea_algorithmic_2015}, stochastic volatility and liquidity \citep{CheriditoSepin2014}, and transient impact through propagator models consistent with no-dynamic-arbitrage conditions \citep{alfonsi2010optimal, Curato2017, Gatheral2010, ObizhaevaWang2013}. More recently, \cite{DupretHainaut2025} derive tractable Hamilton–Jacobi–Bellman (HJB) formulations that link high-frequency microstructure with execution models.

Despite their differences, these models share a common structural feature: market liquidity and price impact are specified exogenously. The trader optimizes against a given impact process (e.g.\ the square-root law), and the resulting problem is a stochastic optimal control problem. However, in modern electronic markets, liquidity is supplied by strategic and adaptive agents. If market makers update their quoting behavior in response to the observed order flow, then price impact is not a fixed structural process but the outcome of interaction. In such a setting, the execution problem is no longer purely a control problem against a passive environment, but it becomes part of a strategic equilibrium problem. Consequently, treating impact as exogenous when liquidity is in fact endogenous constitutes a \textbf{structural misspecification}. Optimality claims derived under fixed impact coefficients need not remain valid once market makers adapt strategically. In particular, a liquidity trader's execution strategy may alter future market states, which in turn affect the subsequent liquidity conditions she faces. This observation motivates the central question of this paper:
\textit{How can optimal execution be studied in a market where liquidity and price formation arise endogenously from strategic interaction among heterogeneous agents?}
Addressing this question requires an environment that is consistent with equilibrium reasoning rather than one that assumes fixed impact dynamics.
\\
\\
Recently, reinforcement learning (RL) has been applied to optimal execution and market making as a data-driven alternative to analytical control methods. Yet most of this literature operates within the same structural paradigm described above: the execution agent is trained in an environment where price impact is either specified parametrically or generated by a simulator with fixed rules. In other words, the underlying liquidity process remains effectively exogenous. Early contributions employed tabular Q-learning methods \citep{hendricks2014reinforcement, nevmyvaka2006reinforcement}. More recent studies adopt deep RL algorithms such as (double) deep Q-learning (DQN), actor–critic methods, and Proximal Policy Optimization (PPO) \citep{cheridito2025reinforcementlearningtradeexecution, espana2025reinforcementlearningqueuereactivemodels,
ganesh2019reinforcement,
hafsi2024optimalexecutionreinforcementlearning, Karpe_2020, Lin2020, Agent-Based-Market-Simulator, Nagy_2023, Ning2018DDQNForOptExecution}. \cite{MoallemiWang2022} formulate execution timing as an optimal stopping problem and demonstrate empirical cost reductions using reinforcement and supervised learning techniques. Deep differentiable RL approaches further incorporate structural model components into policy optimization \citep{Jaisson2022}.
These methods enhance adaptivity and allow for nonlinear trading policies. However, even when learning is fully data-driven, the training environment typically does not model strategic adaptation among market makers. As a result, execution remains formulated as a single-agent stochastic control problem rather than as part of a stochastic game with endogenous liquidity and heterogeneous market players.
\\
\\
In contrast, market microstructure theory models liquidity and price formation as equilibrium outcomes of strategic interaction. In the seminal framework of \cite{kyle_continuous_1985}, prices adjust in response to order flow generated by informed traders interacting with competitive market makers and noise traders. Related models analyze liquidity provision and bid-ask spreads under inventory and adverse-selection effects \citep{glosten_bid_1985, Ho_1981}. \cite{Huberman2004Manipulation} characterize recursive linear equilibria under no-manipulation constraints.  In these models, liquidity depth and impact coefficients are endogenous equilibrium objects rather than exogenous parameters while prices incorporate information through strategic interaction among heterogeneous agents. This theoretical foundation therefore confirms that price formation emerges as a stochastic game rather than as an externally specified process. However, these Kyle-style analytical equilibrium models typically rely on restrictive assumptions, such as linear strategy parameterizations, Gaussian uncertainty, and perfect competition. While these assumptions enable tractable recursive solutions, they limit the models' direct applicability to realistic limit-order-book (LOB) environments and to execution setting involving heterogeneous agents.
\\
\\
Recent machine learning research has explored the numerical approximation of such equilibrium structures. \cite{friedrich2020deep} use adversarial training to recover the static Kyle equilibrium, while \cite{LiEtAl2024} examine the efficiency of market impact estimation from price trajectories. In parallel, multi-agent reinforcement learning (MARL) provides computational tools for modeling strategic interaction among adaptive agents, including independent learning approaches such as IPPO and centralized training decentralized execution (CTDE) methods such as MADDPG \citep{lowe_multi-agent_2017} and its extensions MAPPO \citep{kuba_trust_2022} and HAPPO \citep{yu_surprising_2022}. MARL does not, in general, guarantee convergence to a Nash equilibrium. Nevertheless, if the learning dynamics stabilize and subsequent policy updates cease to induce substantial behavioral changes, the resulting policy profile may be interpreted operationally as an equilibrium-like configuration. In this paper, we adopt this stability-based interpretation without claiming formal convergence to Nash equilibrium by using structural and statistical diagnostics.

However, financial applications rarely integrate such equilibrium-oriented learning with realistic LOB simulation under carefully controlled information structures. A framework that combines strategic learning with realistic market dynamics therefore remains largely underdeveloped.
\\
\\
To study optimal execution under endogenous liquidity, three elements are required.
\begin{enumerate}
    \item First, a mechanism ensuring a well-posed stochastic trading game with synchronized actions and controlled information flow. If agents observe inconsistent states or act on leaked information, the resulting process does not define a valid stochastic game, and learned strategies optimize against a misspecified objective. 
    \item Second, a theoretical framework for approximating equilibrium-like behavior among heterogeneous agents. Such a framework should be validated against a known benchmark equilibrium to demonstrate that it can recover meaningful equilibrium-like structures and price formation. 
    \item Third, a methodology for embedding the liquidity trader's execution problem within this stochastic trading game.
\end{enumerate}

\noindent To achieve each of these three elements, we propose a framework for analyzing optimal execution under endogenous liquidity using MARL within a realistic LOB simulator:
\begin{enumerate}
\item \hspace{-0.5mm}is addressed by constructing a novel synchronized multi-agent market simulator, called \textbf{ABIDES-MARL}, in which heterogeneous agents--including an informed trader, a liquidity trader, competing market makers, and noise traders--interact through a LOB while preserving strict control over information disclosure and action timing.

\item \hspace{-0.5mm}is tackled by validating the framework in a benchmark trading game with asymmetric information inspired by \cite{kyle_continuous_1985}. We show that under appropriate constraints on the policy parameterization, the learning dynamics reproduce the gradual incorporation of private information into prices. When policy classes are overly flexible, agents exploit predictability in systematic ways, leading to persistent deviations from the theoretical benchmark.

\item \hspace{-0.5mm}is achieved by studying optimal execution in the presence of adaptive market makers. Unlike classical exogenous-impact models, execution strategies now both influence and respond to market behavior. The resulting dynamics differ structurally from single-agent benchmarks, demonstrating that equilibrium consistency is essential for valid execution analysis.
\end{enumerate}
To implement this ABIDES-MARL framework, we build on ABIDES (Agent-Based Interactive Discrete Event Simulation) \citep{byrd_abides_2019}, an event-driven multi-agent market simulator, and ABIDES-Gym \citep{amrouni_abides-gym_2021}, which provides a single-agent RL interface. However, ABIDES-Gym supports only one learning agent and cannot enforce synchronized multi-agent decisions without risking information leakage. We therefore introduce architectural modifications that enable equilibrium-consistent multi-agent learning while preserving realistic LOB mechanics.
\\
\\
To summarize, this paper shows that
\textit{a valid analysis of optimal execution under endogenous liquidity requires a market environment that is consistent with equilibrium behavior}. We demonstrate that
MARL can approximate equilibrium-like behavior in a realistic LOB setting with a finite population of heterogeneous market players. When execution is evaluated within such an endogenous environment, the resulting strategies differ structurally from those derived under exogenous price impact assumptions. Our contributions are therefore threefold:
\begin{enumerate}
    \item \textbf{Execution insight:} An optimal execution setting under endogenous liquidity, which is shown to differ fundamentally from solutions derived in exogenous impact models.
    \item \textbf{Equilibrium-approximation framework:} A MARL-based theoretical framework for approximating equilibrium-like behavior in finite-agent trading games.
    \item \textbf{Market simulator design:} ABIDES-MARL, a synchronized multi-agent extension of ABIDES and ABIDES-Gym that enforces controlled information flow and prevents information leakage.
\end{enumerate}
The remainder of the paper is organized as follows. 
Section~\ref{sec:system} introduces the ABIDES-MARL market simulator and formalizes the stochastic trading game. 
Section~\ref{method:MARL-methodology} develops the execution models by transitioning from stochastic optimal control under exogenous liquidity to a stochastic game under endogenous liquidity. 
Section~\ref{sec:numerical-results} presents numerical evidence on equilibrium validation through price discovery and on structurally distinct execution strategies. 
Section~\ref{sec:discussion} concludes. 
All code and configurations are publicly released for reproducibility.\footnote{Repository: \url{https://github.com/10258392511/MARLOptExecution}.}

\section{Market Simulator: ABIDES-MARL}\label{sec:system}

ABIDES is an \textbf{event-driven} multi-agent simulation system designed to model realistic LOB markets \citep{byrd_abides_2019}. RL methods, by contrast, are formulated in \textbf{discrete time}. MARL further requires that all agents observe a consistent market state before acting. This section explains how ABIDES-MARL reconciles these structural differences and ensures that the resulting interaction defines a well-posed stochastic trading game.
\subsection{From discrete event simulation to timestep-based MARL}

ABIDES operates as a discrete-event multi-agent simulation in which agents communicate and submit trading actions through timestamped messages to the exchange. These messages are scheduled and processed sequentially by a central kernel using a priority queue. Because scheduling is entirely timestamp-driven, there is no fixed time grid. State transitions occur asynchronously whenever events are processed by the kernel. Reinforcement learning, in contrast, assumes a timestep-based structure with a regular time grid. At discrete time \(n\), an agent observes a state \(\mathbf{s}_n\), selects an action \(\mathbf{x}_n\), and the environment transitions to a new state \(\mathbf{s}_{n+1}\). In MARL with simultaneous actions, all agents must observe the same market state at time \(n\) before choosing their respective actions. This difference creates a structural incompatibility. ABIDES is inherently asynchronous and event-driven, whereas RL requires a regular time grid to support timestep-based state transitions. Moreover, MARL requires a consistent observation time so that agents do not observe state updates generated by others before selecting their own actions. Without such synchronization, the interaction cannot be interpreted as a well-defined stochastic game. Bridging these paradigms therefore requires controlled interruption of the simulation kernel and a carefully designed synchronization protocol.

\subsection{Kernel synchronization, filtration, and well-posedness}

\begin{figure}
    \centering
    \includegraphics[width=\linewidth]{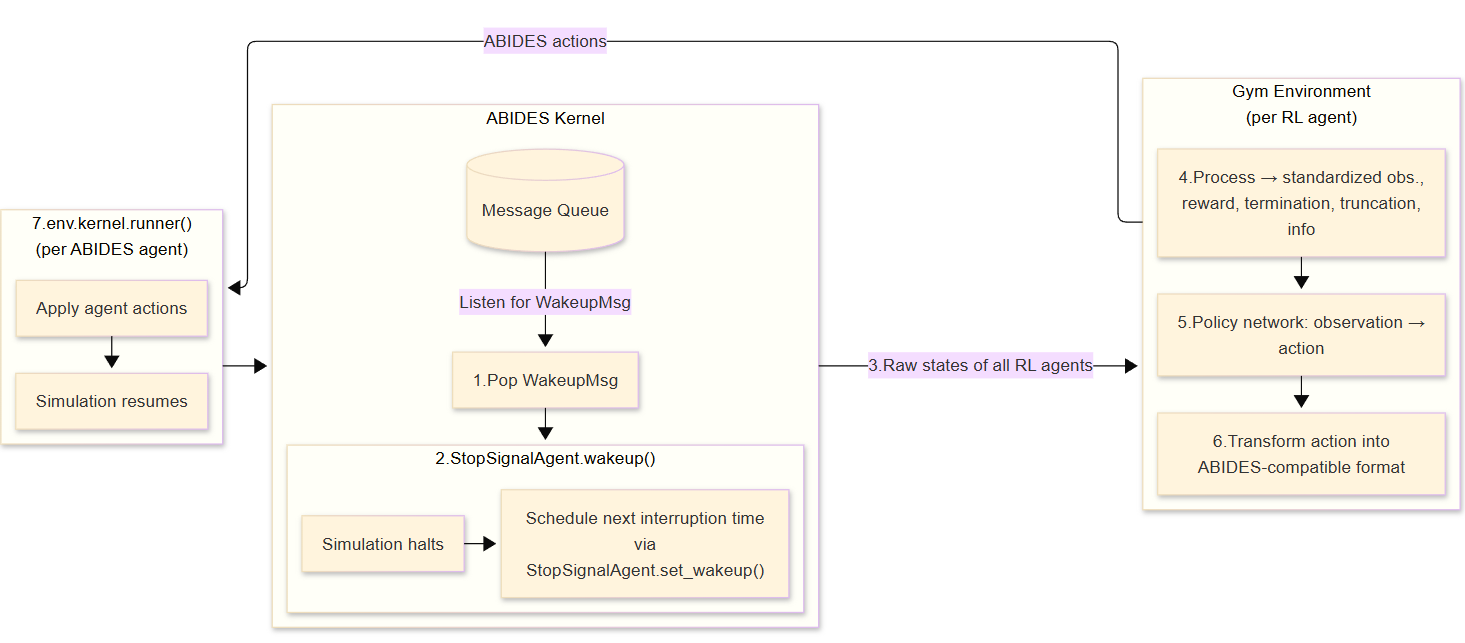}
    \caption[ABIDES-MARL]{ABIDES-MARL communication cycle.}
    \label{fig:ABIDES-MARL}
\end{figure}

ABIDES-Gym partially bridges the gap between event-driven simulation and RL by introducing a special agent type, the \texttt{FinancialGymAgent}. This agent can interrupt the priority-queue-based kernel, collect raw simulation states, compute rewards, and return the standard RL interface tuple consisting of the next observation, reward, termination, truncation, and auxiliary information. However, this design is difficult to extend to a multi-agent setting. If multiple \texttt{FinancialGymAgent}s interrupt the kernel independently, it becomes impossible to guarantee a consistent ordering of actions across learning agents. Sequential trading structures cannot be enforced reliably, and information may be revealed unintentionally if one agent observes a state that already reflects the action of another agent within the same decision period. Such behavior violates the filtration structure required for a well-defined stochastic game. Agents may observe inconsistent states, and their actions may depend on information that should not yet be available. In this case, learned strategies optimize against an artificial objective determined by implementation artifacts rather than by the intended economic model.
\\
\\
To address this issue, ABIDES-MARL introduces a dedicated \texttt{StopSignalAgent}. There is exactly one such agent in the simulation, and only this agent is permitted to interrupt the kernel. At each decision period, the \texttt{StopSignalAgent} pauses the kernel, schedules the next wakeup time, and triggers state collection for all RL agents simultaneously. Raw observations are collected before any learning agent acts. Each agent then computes its action, which is translated into a valid order submission and sent to the exchange. After all actions have been submitted according to a predefined protocol, control is returned to the kernel and the simulation resumes. This design ensures synchronized decision periods and strict control over the ordering of actions.
\\
\\
The synchronization mechanism  also enables a precise filtration structure. Let \(\mathcal{F}_{n-1}\) denote the \(\sigma\)-algebra generated by all publicly observable information up to period \(n-1\). At timestep \(n\), all agents observe the same information set \(\mathcal{F}_{n-1}\), and their actions \(\mathbf{x}_n\) are measurable with respect to \(\mathcal{F}_{n-1}\). The system transitions to the updated information set \(\mathcal{F}_n\) only after all actions at period \(n\) have been executed. In particular, no agent observes any \(\mathcal{F}_n\)-measurable information before choosing its action at time \(n\). Under this construction, the trading interaction satisfies the basic measurability requirements of a stochastic game. Actions are chosen with respect to a common information set, and the state update occurs only after the action profile has been fixed, see Section \ref{sec:multi-agent-RL} for more details. This in turn ensures that the learning problem is well-posed from a probabilistic perspective. Figure~\ref{fig:ABIDES-MARL} illustrates the communication cycle of ABIDES-MARL. The cycle consists of kernel interruption, observation collection, policy inference, action transformation, order submission, and kernel resumption. This controlled procedure ensures that the stochastic game interpretation is enforced at the system level.

\subsection{Trading protocol and information structure}

The synchronized environment allows us to implement a hybrid trading protocol that combines simultaneous and sequential interaction. Agents are divided into two groups. Group A consists of traders, including the informed trader, the liquidity trader, and non-learning noise traders. Group B consists of market makers. Within each group, actions are taken simultaneously, whereas across groups the interaction is sequential. Specifically, at trading period \(n\), the general (high-level) structure of the stage game proceeds as follows\footnote{see Section \ref{sec:multi-agent-RL} for a  detailed description of each step and of each market player.}:
\begin{enumerate}
    \item The informed trader and the liquidity trader choose their order quantities \(x^{\text{IT}, (n)}\) and \(x^{\text{LT}, (n)}\), while noise traders submit independently generated random orders. Without loss of generality, these random orders are aggregated and denoted by \(u^{(n)}\).
    \item Market makers observe the aggregate order flow
    \begin{equation}
        q^{(n)} := x^{\text{IT}, (n)} + x^{\text{LT}, (n)} + u^{(n)}. \label{eq:order-flow}   
    \end{equation}
    \item Each of the \(M\) market makers independently chooses a quote \(p_i^{(n)}\), where \(i = 1, \ldots, M\).
    \item The exchange allocates orders according to quote competitiveness and the market clears at a common price $p^{(n)}$ that is aggregated from the quotes $(p_1^{(n)}, \ldots, p_M^{(n)})$. 
\end{enumerate}
The stage game can therefore be defined as \(G = (A_T, A_{MM}, O, T)\), where \(A_T\) denotes the traders' action space, \(A_{MM}\) denotes the market makers' action space, \(O\) specifies the observation structure, and \(T\) denotes the transition rule. This stage-game representation will be used and detailed in Section~\ref{sec:multi-agent-RL} to formulate optimal execution under endogenous liquidity as a finite-horizon stochastic game.
\\
\\
The information structure $O$ is explicitly controlled and is partitioned into global and local components.

We consider two global observation modes. In \texttt{Exchange} mode, the full LOB is observable to all agents. In over-the-counter (\texttt{OTC}) mode, only the volume-weighted average price is observable, that is, only a summary statistic of the full LOB is revealed. In both modes, all agents observe the current time \(t^{(n)} \in [0, 1]\).

In addition to global information, each agent observes local variables that are not observable to other agent types. The informed trader observes the constant fundamental value \(v\) of the single risky asset. The liquidity trader observes the remaining inventory to be acquired, denoted by \(Q^{(n)}\). Without loss of generality, we consider the liquidity trader performing an acquisition task. Each market maker then observes the aggregate order flow \(q^{(n)}\). A formal specification of the corresponding information sets and admissible strategy spaces for each agent is provided in Section~\ref{sec:multi-agent-RL}.
\\
\\
In summary, ABIDES-MARL defines a synchronized timestep-based stochastic trading game within an event-driven simulation system. It aligns decision periods, controls information flow, and enforces the sequential trading protocol. No agent observes future information before acting. As a result, equilibrium approximation via MARL is well-defined within this environment. We now turn to the formal market model and learning formulation.

\section{Optimal Execution under Exogenous and Endogenous Liquidity}\label{method:MARL-methodology}

This section develops the execution problem in two steps. We first treat in Section \ref{sec:single-agent-RL} market impact as exogenous, which yields a classical finite-horizon stochastic control formulation of the problem for a single liquidity trader. This benchmark admits an explicit dynamic programming solution and serves as a reference point for RL verification. In Section \ref{sec:multi-agent-RL}, we then endogenize liquidity by introducing adaptive market makers whose quoting policies jointly determine the effective market impact. In this setting, price dynamics become policy-dependent, and execution is no longer a single-agent control problem but a finite-horizon stochastic game.

\subsection{Optimal execution with exogenous market impact}\label{sec:single-agent-RL}

\subsubsection{Model formulation}

We consider an acquisition problem over a finite horizon of \(N\) trading periods. The liquidity trader must buy a total quantity \(Q > 0\). Let \(x^{\text{LT}, (n)} \geq 0\) denote the executed quantity at period \(n=1,\ldots,N\), and let \(Q^{(n)}\) denote the remaining inventory to acquire at the beginning of period \(n\). The inventory dynamics are
\[
    Q^{(n)} = Q^{(n - 1)} - x^{\text{LT}, (n-1)}, 
    \qquad Q^{(0)} = Q,
\]
together with the terminal constraint $
    \sum_{n = 1}^N x^{\text{LT}, (n)} = Q,
$ ensuring that the target quantity $Q$ is acquired exactly by the end of the horizon.
Prices are modeled by separating a pre-trade price \(\hat{p}^{(n)}\) with permanent price impact and an execution price \(p^{(n)}\) that includes temporary impact. The pre-trade price evolves according to
\begin{equation*} 
    \hat{p}^{(n)} = \alpha \hat{p}^{(n - 1)} + (1 - \alpha) p^{(n - 1)} + \epsilon^{(n)},
\end{equation*}
where \(\alpha \in [0, 1]\) controls how strongly the last execution price affects future pre-trade levels, serving as a permanent-impact (or mean-reversion) proxy, and \(\epsilon^{(n)} \sim \mathcal{N}(0, \sigma_\epsilon^2)\) is a mean-zero innovation representing a news component. The execution price is then
\[
    p^{(n)} = \hat{p}^{(n)} + \lambda^{(n)}\big(x^{\text{LT}, (n)} + u^{(n)}\big),
\]
where \(u^{(n)} \sim \mathcal{N}(0, \sigma_u^2)\) is mean-zero noise-trader order flow, independent of \(\epsilon^{(n)}\), while \(\lambda^{(n)}\) is an exogenous  deterministic temporary impact process. This benchmark is \emph{exogenous-liquidity} in the sense that the entire process \(\{\lambda^{(n)}\}_{n = 1}^N\) is specified outside the model and does not depend on the trader's behavior or on other strategic agents. In contrast to \cite{almgren_optimal_2001}  which assumes constant market impact parameters, our formulation allows \(\lambda^{(n)}\) to vary over time, consistent with the dynamic microstructure of real markets. Moreover, unlike \cite{Huberman2005Liquidity} which uses the variance of total execution cost as a risk measure (thereby complicating RL reward design), we provide here a tractable formulation suitable for RL. 
\\
\\
The execution problem can then be formulated as the following stochastic control problem:
\begin{align}
    L(Q, N) 
    &:= 
    \min_{\{x^{\text{LT}, (n)} \in \mathbf{M}(H^{(n)})\}_{n = 1}^N}
    \mathbb{E}\left[
        \sum_{n = 1}^N p^{(n)} x^{\text{LT}, (n)} 
        + 
        \phi \sum_{n = 1}^N \big(Q^{(n)}\big)^2
    \right]
    \label{eq:methods-liquidity-opt} \\
    \text{s.t.} \quad 
    & \sum_{n = 1}^N x^{\text{LT}, (n)} = Q, \nonumber
\end{align}
where
\begin{align}
    \hat{p}^{(n)} 
    &= 
    \alpha \hat{p}^{(n-1)} + (1 - \alpha) p^{(n-1)} + \epsilon^{(n)}, 
    \label{eq:methods-liquidity-permanent-price-impact} \\
    p^{(n)} 
    &= 
    \hat{p}^{(n)} + \lambda^{(n)} \big(x^{\text{LT}, (n)} + u^{(n)}\big), 
    \label{eq:methods-liquidity-temp-price-impact} \\
    H^{(n)} 
    &= 
    \left(
        \{ \hat{p}^{(j)} \}_{j = 1}^n, 
        \{ p^{(j)} \}_{j = 1}^{n - 1}, 
        \{ x^{\text{LT}, (j)} \}_{j = 1}^{n - 1},
        \{ u^{(j)} \}_{j = 1}^{n - 1}, 
        \{ \epsilon^{(j)} \}_{j = 1}^n,  Q^{(n)}
    \right),
    \label{eq:methods-liquidity-info-set}
\end{align}
and \(\phi \geq 0\) controls the degree of inventory risk aversion. The notation \(\mathbf{M}(H^{(n)})\) indicates that admissible controls \(x^{\text{LT}, (n)}\) are measurable with respect to the trader’s information set \(H^{(n)}\), capturing that actions are conditioned only on information available up to period \(n\). In particular at period \(n\), the liquidity trader does not observe \(u^{(n)}\) nor the execution price \(p^{(n)}\) when choosing \(x^{\text{LT}, (n)}\). We characterize the optimal strategy of the above optimization problem through a backward difference equation.

\begin{theorem}
\label{thm:liquidity-opt}
Let \( \mu^{(n)} \) satisfy the backward recursion:
\begin{align*}
    \mu^{(n)} 
    &= 
    \alpha \lambda^{(n-1)} + \lambda^{(n)} + \phi 
    - 
    \frac{ (\lambda^{(n)})^2 (1 + \alpha)^2 }{ 4 \mu^{(n+1)} }, 
    \quad \forall n \in \{1, \ldots, N - 1\}, \\
    \mu^{(N)} 
    &= 
    \alpha \lambda^{(N-1)} + \lambda^{(N)} + \phi.
\end{align*}
If \( \mu^{(n)} > 0 \) for all \( n \), then the unique and time-consistent optimal strategy is
\begin{align*}
    x^{\text{LT}, (n)} 
    &= 
    \theta^{(n)} Q^{(n)}, 
    \quad \text{where} \\
    \theta^{(n)} 
    &= 
    1 
    - 
    \frac{ \lambda^{(n)} (1 + \alpha) }{ 2 \mu^{(n+1)} }, 
    \quad \text{for } n < N, 
    \quad \text{and} 
    \quad \theta^{(N)} = 1.
\end{align*}
The corresponding minimal cost at step \( n \) is given by
\[
    L^{(n)}\!\left(\tilde{p}^{(n-1)}, Q^{(n-1)}, Q^{(n)}\right) 
    :=
    \left( 
        \tilde{p}^{(n-1)} 
        - 
        \alpha \lambda^{(n-1)} \big(Q^{(n-1)} + u^{(n-1)}\big) 
    \right) 
    Q^{(n)} 
    + 
    \mu^{(n)} \big(Q^{(n)}\big)^2,
\]
with $\tilde{p}^{(n-1)} := p^{(n-1)} + \epsilon^{(n)}$,  $Q^{(0)} := Q$, 
\( \lambda^{(0)} := 0 \), \( u^{(0)} := 0 \), and $L^{(0)}\!\left(\tilde{p}^{(-1)}, Q^{(-1)}, Q^{(0)}\right) := L(Q,N)$.

\end{theorem}

\paragraph{Remarks}
\begin{itemize}
    \item The optimal strategy depends only on the current inventory \( Q^{(n)} \), and not on the noise order flow \( u^{(n)} \) or news shocks \( \epsilon^{(n)} \), due to their zero-mean property.
    \item The existence of a unique, time-consistent solution is guaranteed under the positivity condition on \( \mu^{(n)} \), which ensures strict convexity of the value function at each backward step.
    \item Although the model could be extended to allow for a stochastic price-impact process \(\lambda^{(n)}\) in the spirit of \cite{CheriditoSepin2014}, we retain a simpler benchmark here for RL comparison.
\end{itemize}

\subsubsection{Single-agent RL formulation}

The above exogenous benchmark can be written as a standard single-agent RL problem. The purpose of this formulation is to verify that RL recovers the analytical solution in the single-agent setting and thereby provides a baseline from which we later construct the endogenous multi-agent formulation. Numerical results are presented in Section~\ref{exp:single-agent}.

\paragraph{State representation}
The observation at step \(n\) is
\[
    \big[t^{(n)},\, p^{(n-1)},\, Q^{(n)}\big],
\]
where \(p^{(n-1)}\) denotes the previous execution price and \(Q^{(n)}\) denotes the remaining inventory to acquire. This reduced state representation is indeed sufficient in our setting. By Theorem~\ref{thm:liquidity-opt}, the minimal cost-to-go at step \(n\) depends only on \(\big[t^{(n)},\, \tilde{p}^{(n-1)},\, Q^{(n)}\big]\) and the chosen action \(x^{\text{LT},(n)}\). Moreover, since the optimal strategy does not depend on the innovation \(\epsilon^{(n)}\) beyond its zero-mean property, we omit it in the RL implementation. Consequently, \(\tilde{p}^{(n-1)}\) can be replaced by \(p^{(n-1)}\), which captures the relevant price information through the previous execution price incorporating temporary impact.

\paragraph{Action space}
At each step \(n\), the agent selects a trading intensity \(\theta^{(n)} \in [0,1]\), corresponding to the order size
\[
    x^{\text{LT},(n)} = \theta^{(n)} Q^{(n)}.
\]
We do not explicitly enforce \(\theta^{(N)} = 1\). Instead, a terminal penalty on unfilled inventory is incorporated into the reward structure to encourage full completion of the target position.

\paragraph{Reward structure}
For \(n \in \{1,\ldots,N-1\}\), the liquidity trader receives the step-wise reward
\begin{equation*}
    R^{\text{LT},(n)}
    =
    -\left(
        x^{\text{LT},(n)} p^{(n)}
        +
        \phi \big(Q^{(n)}\big)^2
    \right).
\end{equation*}
Here, \(p^{(n)}\) is the execution price at the current step, incorporating temporary impact. To incentivize exact acquisition of the target inventory, we add a terminal penalty on any remaining unfilled quantity at step \(N\):
\begin{align*}
    Q_{\text{unfilled}}
    &= 
    Q^{(N)} - x^{\text{LT},(N)}, \\
    R^{\text{LT},(N)}
    &=
    -\left(
        x^{\text{LT},(N)} p^{(N)}
        +
        \phi \big(Q^{(N)}\big)^2
    \right)
    -
    \gamma \big(Q_{\text{unfilled}}\big)^2, 
\end{align*}
where \(\gamma \ge 0\) is chosen substantially larger than \(\phi\). \vspace{-3mm}

\subsection{Optimal execution under endogenous liquidity: the multi-agent trading game}\label{sec:multi-agent-RL}

We now endogenize market impact by explicitly modeling liquidity  as the outcome of strategic interaction among market makers. The resulting structure can be interpreted as an optimal execution problem embedded within a Kyle-type market setting \citep{kyle_continuous_1985}, in which prices are formed through competition among market makers 
in response to order flow. For a summary of the main result in \cite{kyle_continuous_1985}, see  Appendix~\ref{appendix:kyle-solution}. 

In contrast to Section~\ref{sec:single-agent-RL}, where the impact path \(\{\lambda^{(n)}\}_{n = 1}^N\) is specified exogenously, here the effective impact coefficient at each step is an endogenous function of market makers' actions. Consequently, the liquidity trader's problem is no longer a single-agent stochastic control problem: the state dynamics depend on the joint policy profile of multiple learning agents, and the environment must be treated as a finite-horizon stochastic game. Formally, this stochastic game is obtained by repeating over \(n = 1, \ldots, N\) the stage game \(G = (A_T, A_{MM}, O, T)\) introduced in Section~\ref{sec:system}, with state transitions and payoffs determined by the joint policy profile. We next describe this extended Kyle-style execution trading game in the context of endogenous liquidity provision.

\paragraph{Agents and horizon}
Fix a finite horizon \(N \in \mathbb{N}\). The set of learning agents is
\[
    \mathcal{A} = \{\text{IT}, \text{LT}, \text{MM}_1, \ldots, \text{MM}_M\},
\]
consisting of one informed trader (IT), one liquidity trader (LT), and \(M\) competing market makers (MMs). In addition, a representative noise trader generates random order flow but does not learn. We now detail the structure of the stage game \(G = (A_T, A_{MM}, O, T)\) introduced in Section~\ref{sec:system}.
\paragraph{Initial information} We denote $v \sim \mathcal{N}(p_0, \sigma^2_0)$, the \textit{ex post} fundamental (liquidation) value of the risky asset. Before trading begins, the exogenous value
of $v$ is realized and is observed only by the informed trader.
\paragraph{\hspace{-0.4mm}Extended \hspace{-0.2mm}Kyle-style \hspace{-0.2mm}execution \hspace{-0.2mm}trading game}\hspace{-0.4mm}At \hspace{-0.2mm}each \hspace{-0.2mm}trading \hspace{-0.2mm}round $n=1,\ldots,N$, sequentially,
\begin{enumerate}
    \item[\textbf{(i)}] Simultaneously,\vspace{1mm}
    \begin{itemize}
        \item[-] \textit{Noise traders} submit \(u^{(n)} \sim \mathcal{N}(0,\sigma^2_u)\), an exogenous mean-zero noise order flow, assumed i.i.d. across \(n\) and from $v$.\vspace{1mm}
        \item[-]  The \textit{liquidity trader} submits an acquisition order \(x^{\text{LT}, (n)} \in \mathbb{R}_{+}\) based on the remaining inventory $Q^{(n)}$ and on the public history of execution prices  $\{p^{(1)},\dots,p^{(n-1)}\}$ (or LOB history  $\{\mathrm{LOB}^{(j)}\}_{j=1}^{n-1}$, see Definition \ref{method:def-LOB} below). \vspace{1mm}
        \item[-]  The \textit{informed trader} submits an order \(x^{\text{IT}, (n)} \in \mathbb{R}\) based on $v$ and the public history of execution prices  $\{p^{(1)},\dots,p^{(n-1)}\}$ (or LOB history  $\{\mathrm{LOB}^{(j)}\}_{j=1}^{n-1}$).
    \end{itemize}
    \vspace{2mm}
    \item[\textbf{(ii)}] Each \textit{market maker} $i=\{1,\ldots, M \}$ then observes the the public history of execution prices  $\{p^{(1)},\dots,p^{(n-1)}\}$ (or LOB history  $\{\mathrm{LOB}^{(j)}\}_{j=1}^{n-1}$) and the aggregate signed order flow \vspace{-2mm}
    \[
    q^{(n)} = x^{\text{IT}, (n)} + x^{\text{LT}, (n)} + u^{(n)}. \vspace{-2mm}
\] Each market maker $i$ posts a quote \(p_i^{(n)} \in \mathbb{R}_+\), or equivalently\footnote{Note that even if the quoting rule is nonlinear (the MM$_i$ quotes $p^{(n)}_i$ instead of $\lambda^{(n)}_i$), the posted price \(p_i^{(n)}\) in \eqref{eq:linear-mkt-impact} always admits an implied linear impact coefficient $
   \lambda_i^{(n)} 
    =  (p_i^{(n)} - \hat{p}^{(n - 1)})/ q^{(n)},$ for $ q^{(n)} \neq 0$. } specifies a liquidity impact parameter \(\lambda_i^{(n)}\) that determines \(p_i^{(n)}\). Under a linear quoting rule,\vspace{-1mm}
\begin{equation}
    p_i^{(n)} = \hat{p}^{(n - 1)} + \lambda_i^{(n)} q^{(n)}, \label{eq:linear-mkt-impact} \vspace{-1mm}
\end{equation}
where \(\hat{p}^{(n - 1)}\) again denotes the pre-trade price with permanent price impact  \(\alpha \in [0,1]\), \vspace{-2mm}
\begin{equation*}
    \hat{p}^{(n)}
    =
    \alpha \hat{p}^{(n-1)}
    +
    (1-\alpha)p^{(n-1)}. \vspace{-1.2mm}
\end{equation*}
\item[\textbf{(iii)}] Each \textit{market maker} is assigned a fraction of the total order flow $q^{(n)}$, and the market clears at the next common public execution price $p^{(n)}=f(p_1^{(n)}, \ldots ,p^{(n)}_M)$, where $f$ is an aggregation rule applied to the $M$ individual quotes. \vspace{1mm}
\end{enumerate}

\paragraph{LOB, allocation rule and market clearing price} What is left to defined is (iii): the rule for allocating order flow across market makers, as well as the aggregation procedure of the individual quotes $p^{(n)}_i$ into the single public price $p^{(n)}$. This requires a suitable definition of the LOB.

\begin{definition}[Reduced-form LOB]
\label{method:def-LOB}
    Let \(\lambda_i^{(n)}\) denote the (implied) market impact parameter of market maker \(i\) at time step \(n\).  
    The market depth provided by market maker \(i\) is defined as \(d_i^{(n)} := 1/|\lambda_i^{(n)}|\).  
    Let \(i_1^{(n)}, \ldots, i_M^{(n)}\) be the permutation that sorts the quotes \(\{p_i^{(n)}\}_{i=1}^M\) in ascending order.  
    The limit order book at step \(n\) is represented by
    \begin{align}
        \mathrm{LOB}^{(n)} 
        := 
        \begin{bmatrix}
            d_{i_1^{(n)}}^{(n)} & p_{i_1^{(n)}}^{(n)} \\
            \vdots & \vdots \\
            d_{i_M^{(n)}}^{(n)} & p_{i_M^{(n)}}^{(n)}
        \end{bmatrix}.
    \end{align}
\end{definition}

\noindent Specifically, only the sorted depth–price pairs are retained in this representation; the identity of the quoting market maker is not observable from the LOB snapshot. This reduced-form construction reflects the fact that, in the present execution setting, price formation is modeled as an auction-style clearing mechanism rather than a price–time priority queue. Execution prices depend on the competitiveness and aggregate depth of quotes, not on queue positions. This abstraction serves two purposes. First, it aligns the market-clearing mechanism with the pro-rata allocation rule introduced below and in \cite{kyle_continuous_1985}, ensuring a unique execution price and tractable aggregation of liquidity. Second, it provides a parsimonious representation of endogenous impact while remaining compatible with the full limit-order-book infrastructure of ABIDES-MARL. In particular, although we adopt this reduced-form LOB for analytical transparency and to recover known equilibrium characteristics in benchmark cases, the underlying ABIDES-MARL market simulator naturally supports standard price–time priority LOB mechanics, and the framework generalizes directly by modifying the allocation and sorting rules. 
\\
\\
We now use a pro-rata allocation to assign the total order flow $q^{(n)}$ to each market maker. Let $
    d_i^{(n)} := 1/|\lambda_i^{(n)}|$
denote the depth implied by market maker \(i\)'s quote at period \(n\).  
Under pro-rata allocation, the volume assigned to market maker \(i\) is
\begin{equation} \label{eq:pro-rata-order}
    \mathrm{Order}_i^{(n)}
    =
    q^{(n)} \cdot
    \frac{d_i^{(n)}}{\sum_{j = 1}^M d_j^{(n)}}.
\end{equation}
The public execution price is then defined as the volume-weighted average price (VWAP):
\begin{equation}
    p^{(n)}
    :=
    \frac{\sum_{i = 1}^M \mathrm{Order}_i^{(n)} p_i^{(n)}}
         {\sum_{i = 1}^M \mathrm{Order}_i^{(n)}}.
    \label{eq:VWAP-def}
\end{equation}
\begin{proposition}[Unanimous VWAP]
\label{method:claim-unanimous-VWAP}
Under pro-rata allocation, all trades in period \(n\) clear at the common execution price \(p^{(n)}\), which can be rewritten as
\begin{align} \label{price_fin}
    p^{(n)}
    =
    \frac{\sum_{i=1}^M d_i^{(n)} p_i^{(n)}}
         {\sum_{i=1}^M d_i^{(n)}}.
\end{align}
Moreover, under the linear quoting rule \eqref{eq:linear-mkt-impact} for each market maker, the VWAP admits the reduced-form update
\begin{equation}
    p^{(n)}
    =
    \hat{p}^{(n-1)}
    +
    \lambda_{\mathrm{eff}}^{(n)} q^{(n)} ,
\end{equation}
where $\lambda_{\mathrm{eff}}^{(n)} $ is the effective impact coefficient given by
\begin{equation}
    \lambda_{\mathrm{eff}}^{(n)}
    =
    \frac{\sum_{i = 1}^M \lambda_i^{(n)} d_i^{(n)}}
         {\sum_{i = 1}^M d_i^{(n)}} 
    =
    \frac{\sum_{i = 1}^M \mathrm{sign}\!\left(\lambda_i^{(n)}\right)}
         {\sum_{i = 1}^M \frac{1}{|\lambda_i^{(n)}|}} \, .
\end{equation}
\end{proposition}

\begin{proof}
See Appendix~\ref{appendix:proof-unanimous-VWAP}.
\end{proof}
\noindent
Hence, \eqref{price_fin} shows that the execution price is the volume-weighted average of quoted prices, with weights given by quoted depths. Moreover, the effective \(\lambda_{\mathrm{eff}}^{(n)}\) is a depth-weighted aggregation of individual impact parameters and is fully determined by the market makers' period-\(n\) policy outputs. Finally, we specify the objective function of each agent, as required for our MARL implementation.

\paragraph{Objectives}

Each learning agent optimizes a finite-horizon conditional expected objective, subject to its own information constraints. Admissible strategies are required to be measurable with respect to the corresponding agent-specific information sets defined below. This yields a well-posed stochastic game under both \texttt{Exchange} and \texttt{OTC} observation modes.

\medskip

\noindent
\textbf{Liquidity trader.}
The liquidity trader solves
\begin{align}
\label{eq:LT-objective}
    \min_{\{x^{\text{LT},(n)} \in \mathbf{M}(H^{\text{LT}, (n)})\}_{n = 1}^N}
    \mathbb{E}
    \left[
        \sum_{n = 1}^N 
        p^{(n)} x^{\text{LT},(n)}
        +
        \phi \sum_{n = 1}^N \big(Q^{(n)}\big)^2
    \right]
\end{align}
subject to
\begin{align*}
    Q^{(n + 1)} = Q^{(n)} - x^{\text{LT},(n)} \, ,
\end{align*}
with $Q^{(0)} = Q\, \text{ and } \sum_{n = 1}^N x^{\text{LT},(n)} = Q$. 
The information set \(H^{\text{LT}, (n)}\) depends on the observation mode:
\begin{equation*}
H^{\text{LT}, (n)} 
\begin{cases}
\left(
   \{\mathrm{LOB}^{(j)}\}_{j = 1}^{n - 1},\;
   \{x^{\text{LT}, (j)}\}_{j = 1}^{n - 1},\;
   Q^{(n)}
   \right), & \texttt{Exchange}, \\[0.5em]
   \left(
   \{p^{(j)}\}_{j = 1}^{n - 1},\;
   \{x^{\text{LT}, (j)}\}_{j = 1}^{n - 1},\;
    Q^{(n)}
   \right),  & \texttt{OTC}.
   \end{cases}
\end{equation*}
In particular, at period \(n\), the liquidity trader does not observe the contemporaneous aggregate order flow \(q^{(n)}\), the noise order flow \(u^{(n)}\), or the current execution price \(p^{(n)}\) (nor \(\mathrm{LOB}^{(n)}\)) when choosing \(x^{\text{LT}, (n)}\).

\medskip

\noindent
\textbf{Informed trader.}
The informed trader solves
\begin{equation}
\label{eq:IT-objective}
    \max_{\{x^{\text{IT},(n)} \in \mathbf{M}(H^{\text{IT}, (n)})\}_{n = 1}^N}
    \mathbb{E}
    \left[
        \sum_{n = 1}^N
        x^{\text{IT},(n)} \big(v - p^{(n)}\big)
    \right].
\end{equation}
Its information set is
\begin{equation*}
H^{\text{IT}, (n)} =
\begin{cases}
\left(
\{\mathrm{LOB}^{(j)}\}_{j = 1}^{n - 1},\;
\{x^{\text{IT}, (j)}\}_{j = 1}^{n - 1},\;
v
\right), & \texttt{Exchange}, \\[0.5em]
\left(
\{p^{(j)}\}_{j = 1}^{n - 1},\;
\{x^{\text{IT}, (j)}\}_{j = 1}^{n - 1},\;
v
\right), & \texttt{OTC}.
\end{cases}
\end{equation*}
In particular, the informed trader conditions on the public price history together with the privately observed fundamental value \(v\), but does not observe the contemporaneous aggregate order flow \(q^{(n)}\), the noise order flow \(u^{(n)}\), or the current execution price \(p^{(n)}\) (nor \(\mathrm{LOB}^{(n)}\)) when choosing \(x^{\text{IT}, (n)}\).

\medskip

\noindent
\textbf{Market maker \(i\).}
Market maker \(i\) solves
\begin{equation}
\label{eq:MM-objective}
    \max_{\{p_i^{(n)} \in \mathbf{M}(H_i^{\text{MM}, (n)})\}_{n = 1}^N}
    \mathbb{E}
    \left[
        \sum_{n = 1}^N 
        \mathrm{Order}_i^{(n)}
        \big(
            p_i^{(n)} - p^{(n)}
        \big)
    \right].
\end{equation}
Its information set is
\begin{equation*}
H_i^{\text{MM}, (n)} =
\begin{cases}
\left(
\{\mathrm{LOB}^{(j)}\}_{j = 1}^{n - 1},\;
\{p_i^{(j)}\}_{j = 1}^{n - 1},\;
\{q^{(j)}\}_{j = 1}^{n}
\right), & \texttt{Exchange}, \\[0.5em]
\left(
\{p^{(j)}\}_{j = 1}^{n - 1},\;
\{p_i^{(j)}\}_{j = 1}^{n - 1},\;
\{q^{(j)}\}_{j = 1}^{n}
\right), & \texttt{OTC}.
\end{cases}
\end{equation*}
In particular, each market maker observes the contemporaneous aggregate order flow \(q^{(n)}\) before quoting at period \(n\), consistent with the sequential protocol introduced in Section~\ref{sec:system}.
\\
\\
While the above formulation specifies the full admissible information sets required for a well-posed stochastic game, the MARL implementation uses reduced observations rather than the full histories. At period \(n\), the liquidity trader observes \([t^{(n)}, \mathrm{LOB}^{(n - 1)}, Q^{(n)}]\) in \texttt{Exchange} mode and \([t^{(n)}, p^{(n - 1)}, Q^{(n)}]\) in \texttt{OTC} mode. The informed trader observes \([t^{(n)}, \mathrm{LOB}^{(n - 1)}, v]\) in \texttt{Exchange} mode and \([t^{(n)}, p^{(n - 1)}, v]\) in \texttt{OTC} mode. Each market maker \(i\) observes \([t^{(n)}, \mathrm{LOB}^{(n - 1)}, q^{(n)}]\) in \texttt{Exchange} mode and \([t^{(n)}, p^{(n - 1)}, q^{(n)}]\) in \texttt{OTC} mode. This reduction is motivated by benchmark results. In the single-agent setting, Theorem~\ref{thm:liquidity-opt} shows that the optimal policy depends only on the current inventory and the previous price. In the Kyle model \citep{kyle_continuous_1985}, the equilibrium strategies likewise depend only on current sufficient statistics. Although such sufficient conditions cannot be established in general for the full multi-agent environment considered here, this reduced observation design enables direct comparison with both the single-agent benchmark (Section~\ref{sec:single-agent-RL}) and the Kyle benchmark (Appendix~\ref{appendix:kyle-solution}).

\paragraph{Linear versus nonlinear parameterization}

To facilitate comparison with the benchmark linear recursive equilibria given in Appendix~\ref{appendix:kyle-solution}, we consider a linear parameterization. For the informed trader, this takes the form
\begin{equation}
\label{eq:IT-linear-param}
    x^{\text{IT},(n)}
    =
    \beta^{(n)}
    \big(
        v - \hat{p}^{(n-1)}
    \big),
\end{equation}
where \(\beta^{(n)} \in \mathbb{R}\) is chosen by the policy.
For market maker \(i\), the linear quoting rule is
\begin{equation}
\label{eq:MM-linear-param}
    p_i^{(n)}
    =
    \hat{p}^{(n-1)}
    +
    \lambda_i^{(n)} q^{(n)},
\end{equation}
where \(\lambda_i^{(n)} \in \mathbb{R}\). We refer to 
\eqref{eq:IT-linear-param}–\eqref{eq:MM-linear-param}
as \emph{linear parameterization}. 
Under \emph{nonlinear parameterization}, 
the informed trader directly selects \(x^{\text{IT},(n)}\) 
and market maker \(i\) directly selects \(p_i^{(n)}\) 
without imposing linear structure. 
The two schemes are compared numerically in 
Section~\ref{sec:numerical-results}.

\paragraph{Zero profit across market makers}

Under pro-rata allocation and the VWAP definition, 
competition among market makers implies zero aggregate profit at each period. The period-\(n\) reward for market maker \(i\) is
\begin{equation}
\label{eq:seq-kyle-reward-mm}
    R_i^{\text{MM},(n)}
    =
    \mathrm{Order}_i^{(n)}
    \big(
        p_i^{(n)} - p^{(n)}
    \big).
\end{equation}

\begin{lemma}[Zero profit among market makers]
\label{method:lemma-zero-sum-game}
For every period \(n\),
\[
    \sum_{i=1}^M R_i^{\text{MM},(n)} = 0.
\]
\end{lemma}

\begin{proof}
See Appendix~\ref{appendix:proof-zero-sum-game}.
\end{proof}\noindent This effectively shows that our model avoids solving a repeated game at each trading step by relaxing the assumption of perfect competition in \cite{kyle_continuous_1985}. We instead introduce a weaker form of competition for computational tractability, while retaining the essential game-theoretic structure of the sequential Kyle framework.
\paragraph{Endogenous-liquidity execution as a stochastic game}

Because the execution price \(p^{(n)}\) depends on the effective impact coefficient 
\(\lambda_{\mathrm{eff}}^{(n)}\), and because 
\(\lambda_{\mathrm{eff}}^{(n)}\) is a deterministic function of the market makers' period-\(n\) actions, 
the state transition of the liquidity trader depends on the joint actions of the informed trader, liquidity trader and market makers. At the same time, each market maker’s optimal quoting decision depends on anticipated order flow, which itself depends on the liquidity trader’s and informed trader’s actions.  Hence the evolution of prices and inventories is jointly action-dependent. 
Formally, the execution problem under endogenous liquidity is not a stochastic control problem for a single agent, but a finite-horizon stochastic game. The MARL formulation of this game is now described.

\subsection{MARL formulation and equilibrium-like behavior}

This subsection specifies (i) the learning dynamics used to train policies and (ii) the precise sense in which the resulting trained profile is interpreted as “equilibrium-like.” 
We retain a fixed-point perspective as a \emph{diagnostic of convergence of the learning procedure}, while carefully distinguishing it from a Nash equilibrium of the underlying stochastic game.

\subsubsection{Policies, objectives and joint training distribution}

For each agent \(i \in \mathcal{A}\), let 
\(
\pi_i(\cdot; \boldsymbol{\theta}_i)
\)
be a parameterized stochastic policy mapping observations to actions. 
Denote the joint parameter vector by
\[
    \boldsymbol{\theta}
    =
    \big(
        \boldsymbol{\theta}_{\text{IT}},
        \boldsymbol{\theta}_{\text{LT}},
        \boldsymbol{\theta}_{\text{MM}_1},
        \ldots,
        \boldsymbol{\theta}_{\text{MM}_M}
    \big),
\]
and the corresponding joint policy profile by
\[
    \pi_{\boldsymbol{\theta}}
    =
    \big(
        \pi_{\text{IT}}(\cdot; \boldsymbol{\theta}_{\text{IT}}),
        \pi_{\text{LT}}(\cdot; \boldsymbol{\theta}_{\text{LT}}),
        \pi_{\text{MM}_1}(\cdot; \boldsymbol{\theta}_{\text{MM}_1}),
        \ldots,
        \pi_{\text{MM}_M}(\cdot; \boldsymbol{\theta}_{\text{MM}_M})
    \big).
\]
For each agent \(i\), define the objective functional
\[
    J_i(\boldsymbol{\theta})
    :=
    \mathbb{E}^{\pi_{\boldsymbol{\theta}}}
    \left[
        \sum_{n = 1}^N R_i^{(n)}
    \right],
\]
where \(R_i^{(n)}\) denotes the period-\(n\) reward of agent \(i\) and the expectation is taken with respect to the trajectory distribution induced by the environment dynamics under \(\pi_{\boldsymbol{\theta}}\), including randomness from noise order flow \(u^{(n)}\) and policy stochasticity.

\subsubsection{Independent PPO updates and the learning operator}

Training is conducted via independent PPO updates in a shared environment, where each agent optimizes its own objective while treating the other agents as part of the environment. This corresponds to a decentralized learning scheme without a centralized critic. One training iteration consists of:

\begin{enumerate}
    \item \textbf{Rollout:} Sample trajectories under the current joint profile \(\pi_{\boldsymbol{\theta}^{(k)}}\).
    \item \textbf{Update:} For each agent \(i\), apply one PPO improvement step, i.e., using its own value function, advantage estimates, and clipping constraints, to obtain updated parameters
    \[
        \boldsymbol{\theta}_i^{(k+1)}
        =
        \mathcal{U}_i\big(\boldsymbol{\theta}^{(k)}\big),
    \]
    where \(\mathcal{U}_i\) denotes the PPO update map induced by sampled trajectories and the optimization procedure; see \cite{schulman_proximal_2017} for the PPO algorithm and \cite{yu_surprising_2022} for its multi-agent extension.
\end{enumerate}
Collecting all agents, define the joint update map (the \emph{learning operator})
\begin{align}
    \boldsymbol{\theta}^{(k+1)}
    &= 
    \mathcal{L}\big(\boldsymbol{\theta}^{(k)}\big), \\
    \mathcal{L}(\boldsymbol{\theta})
    &:=
    \big(
        \mathcal{U}_{\text{IT}}(\boldsymbol{\theta}),
        \mathcal{U}_{\text{LT}}(\boldsymbol{\theta}),
        \mathcal{U}_{\text{MM}_1}(\boldsymbol{\theta}),
        \ldots,
        \mathcal{U}_{\text{MM}_M}(\boldsymbol{\theta})
    \big).
\end{align}
This defines a discrete-time dynamical system on the parameter space. 
The operator \(\mathcal{L}\) is algorithm-dependent and stochastic due to sampling and finite optimization.

\paragraph{Stationarity as an algorithmic notion}

If training stabilizes empirically, i.e., parameter updates become negligible, then the observed parameter vector \(\boldsymbol{\theta}^*\) is approximately stationary under the learning dynamics:
\[
    \boldsymbol{\theta}^*
    \approx
    \mathcal{L}(\boldsymbol{\theta}^*).
\]
For single-agent PPO, \cite{jin2023stationary} prove stationary-point convergence for PPO-Clip, and \cite{huang2024ppo} give the first global convergence results for a PPO-Clip variant in tabular and neural-function-approximation settings. However, for IPPO, there is no known general contraction/fixed-point convergence theorem in general Markov games for $\mathcal{L}$, see \cite{su2022decentralized}. 
Therefore, this “fixed-point” statement refers to \emph{stationarity of the learning algorithm}, not to equilibrium existence of the underlying stochastic game. 
Specifically, it means that one additional PPO iteration, performed against the same opponents, produces negligible change in policy parameters.

This should also be distinguished from a Nash equilibrium. 
A joint policy profile \(\pi^*\) is a Nash equilibrium if, for every agent \(i\),
\[
    J_i(\pi_i^*, \pi_{-i}^*)
    \ge
    J_i(\pi_i, \pi_{-i}^*)
\quad
\text{for all admissible } \pi_i.
\]
In contrast, 
\(
\boldsymbol{\theta}^* \approx \mathcal{L}(\boldsymbol{\theta}^*)
\)
asserts only that no further improvement is found by one PPO update within the chosen parameterization and optimization procedure. 
The two notions differ in several fundamental respects:

\begin{enumerate}
    \item \textbf{Algorithmic vs.\ game-theoretic optimality.} 
    PPO is not a best-response oracle. 
    Absence of PPO improvement does not imply absence of profitable deviations in the underlying game.
    
    \item \textbf{Restricted deviations.} 
    PPO searches within a parameterized policy class under trust-region-type constraints and finite optimization steps. 
    A Nash equilibrium is defined with respect to all admissible unilateral deviations (or at least all deviations within the specified policy class), not only those reachable via a local PPO update.
    
    \item \textbf{Sampling and state distribution.} 
    The operator \(\mathcal{L}\) depends on trajectories sampled under \(\pi_{\boldsymbol{\theta}^*}\). 
    A profitable deviation may exist but remain undetected if relevant states are rarely visited under the current joint policy.
\end{enumerate}
Accordingly, 
we use here an operational definition of equilibrium-like behavior, but do not claim existence of a Nash equilibrium. We  refer to a trained profile \(\pi_{\boldsymbol{\theta}^*}\) as \emph{equilibrium-like} if it satisfies the \textbf{training stability criterion}: Across independent random seeds, training produces joint profiles that generate similar market statistics and similar agent returns \(J_i\). This operational notion hence provides a practical interpretation of equilibrium-like behavior emerging from the MARL training procedure.

\subsection{Special cases}

\paragraph{Extended Kyle benchmark (no liquidity trader)}

If the liquidity trader is removed, then 
\(
\mathcal{A} = \{\text{IT}, \text{MM}_1, \ldots, \text{MM}_M\}.
\)
Order flow reduces to informed trading plus noise trading, and market makers' quotes respond solely to informational order imbalance. 
This setting corresponds to the classical Kyle framework \citep{kyle_continuous_1985}. This case serves as a benchmark because the classical Kyle model admits a closed-form linear recursive equilibrium under strong structural assumptions. 
Within our framework, when the policy class is restricted to linear parameterizations, the resulting equilibrium characterization can be directly compared to the analytical recursive solution. 
The closed-form linear recursive equilibrium is summarized in Appendix~\ref{appendix:kyle-solution}.  In Section~\ref{exp:Kyle}, we use this benchmark to validate the numerical learning procedure by verifying that it reproduces the known qualitative features of price discovery in a setting where the equilibrium structure is analytically characterized.

\paragraph{Liquidity trader versus market makers (no informed trader)}

If the informed trader is removed, then 
\(
\mathcal{A} = \{\text{LT}, \text{MM}_1, \ldots, \text{MM}_M\}.
\)
In this case, there is no informational asymmetry in favor of the trading side. 
The liquidity trader must execute a predictable target inventory against adaptive market makers, while market makers observe aggregate order flow and face no adverse-selection risk. Under endogenous liquidity, this creates an informational imbalance: market makers anticipate forced execution and may optimally adjust quotes toward extreme price levels. 
As a result, price dynamics may become degenerate, with execution prices reflecting strategic extraction rather than meaningful liquidity provision.

This case isolates the role of informational balance in sustaining non-trivial market dynamics. 
In Section~\ref{exp:LT-vs-MMs}, we show numerically that, in the absence of an informed counterparty, endogenous-liquidity execution may collapse into boundary behavior. This observation also highlights a limitation of exogenous-liquidity models. 
Because such models abstract from informational structure and strategic interaction, they cannot capture the destabilizing effects of informational imbalance. 
As a consequence, optimal policies derived under exogenous impact assumptions may fail to remain meaningful once liquidity becomes endogenous.

\section{Numerical Results}\label{sec:numerical-results}

This section presents numerical evidence supporting the theoretical framework developed in Section~\ref{method:MARL-methodology}. We proceed in two steps. First, we verify that RL recovers the analytical solution in the benchmark execution problem with exogenous liquidity (Section~\ref{sec:single-agent-RL}). This serves as a control experiment, confirming that the learning algorithm correctly solves the classical stochastic control problem under fixed market impact. Second, we study execution in the multi-agent environment where liquidity is generated endogenously by strategic market makers (Section~\ref{sec:multi-agent-RL}). We begin by validating equilibrium-like price discovery in the extended Kyle benchmark, and then examine optimal execution under endogenous liquidity across different policy parameterizations.

\subsection{Single-agent benchmark: exogenous liquidity}

We begin with the benchmark execution problem in which the liquidity coefficient is fixed and does not depend on other agents' behavior. In this setting, execution reduces to a finite-horizon stochastic control problem with known and stationary dynamics.

\paragraph{Environment configuration}
We consider a time-constant impact process
\[
    \lambda^{(n)} \equiv \lambda = 2,
\]
with horizon \(N = 20\), target inventory \(Q = 1000\), running inventory penalty \(\phi = 0.01\), terminal penalty \(\gamma = 100\), and no mean reversion (\(\alpha = 0\)). The initial price is set to \(p^{(0)} = 1000\) cents. Noise order flow \(u^{(n)}\) is i.i.d.\ Gaussian with distribution \(\mathcal{N}(0, \sigma_u^2)\) and \(\sigma_u = 50\). The optimal execution schedule is characterized by the backward recursion in Theorem~\ref{thm:liquidity-opt}.

\paragraph{RL versus analytical solution}
We train a single PPO agent in this fixed-impact environment and compare the learned policy with the analytical benchmark. For validation purposes, the environment implements the price dynamics directly, without incorporating the LOB mechanisms of ABIDES-MARL, such as tick-size constraints and order matching. Training is conducted over 1000 episodes, and evaluation is performed on 30 holdout episodes. The policy network consists of two hidden layers (64 units each) with Tanh activations.

\begin{figure}[h!]
    \centering
    \includegraphics[width=\linewidth]{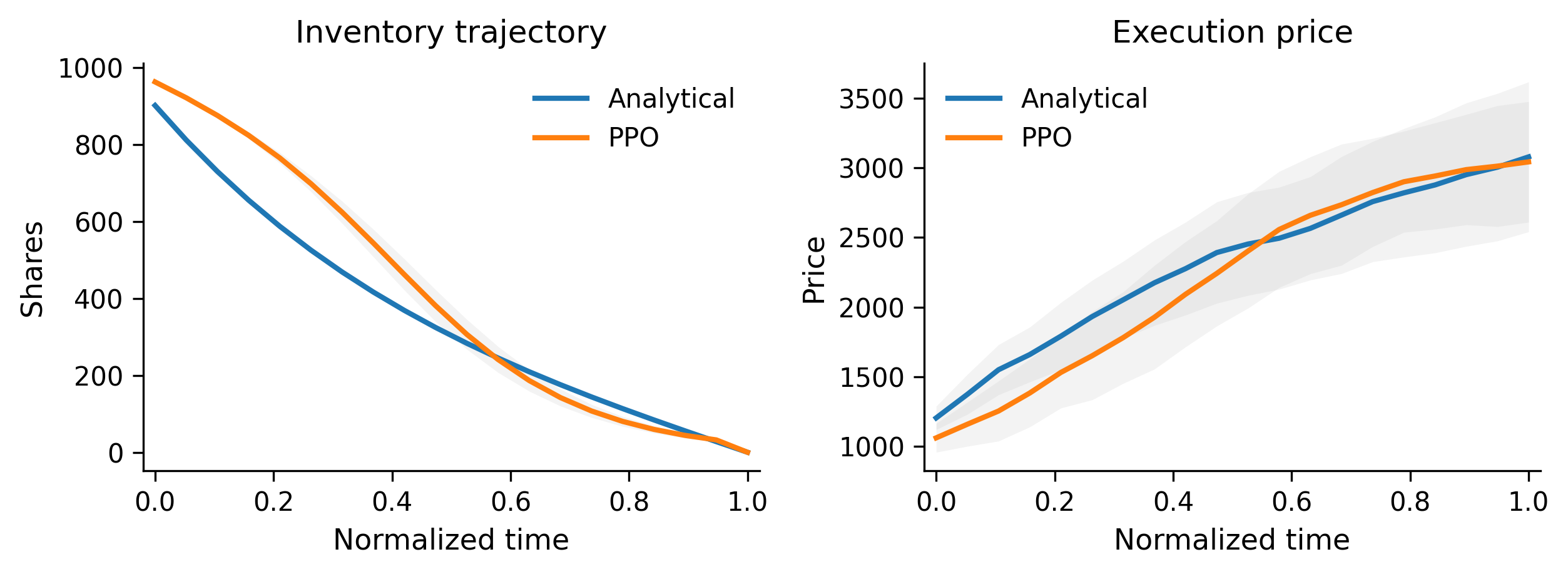}
    \caption{The left panel displays the remaining inventory trajectory \(Q^{(n)}\) under the analytical and PPO policies. The right panel shows the induced transaction price path \(p^{(n)}\).
    \label{fig:single-agent-state-trajectories}}
\end{figure}

\noindent Figure~\ref{fig:single-agent-state-trajectories} shows that the learned PPO policy closely matches the analytical inventory schedule and produces nearly identical price dynamics throughout the horizon. This experiment provides a numerical sanity check. In a stationary environment with fixed impact, RL successfully recovers the dynamic-programming solution. Consequently, any deviations observed in the endogenous-liquidity experiments below cannot be attributed to a failure of the learning algorithm to solve the baseline control problem. Rather, such deviations arise because, once liquidity is generated by strategic agents, the effective impact becomes policy-dependent and the execution problem is no longer governed by fixed coefficients.

\subsection{Multi-agent experiments: endogenous liquidity environment}\label{exp:single-agent}

We now turn to the full stochastic game in which liquidity is generated endogenously by strategic market makers. In this environment, the effective market impact depends on the joint policy profile, and execution becomes policy-dependent.

\paragraph{Experimental setup}

All experiments are conducted in the synchronized ABIDES-MARL environment with horizon \(N = 20\). Unless stated otherwise, the fundamental value \(v\) is drawn independently at the start of each training episode from
\[
    v \sim \mathcal{N}(\mu_v = 1000, \sigma_v^2 = 100^2),
\]
and prices are constrained to the symmetric band \([p_{\min}, p_{\max}] = [0.5v, 1.5v]\) to ensure numerical stability. During evaluation, we fix \(v = \mu_v\) and consider, without loss of generality, two initial conditions for the opening price: \(p^{(0)} = 0.7 \mu_v\) (``down'') and \(p^{(0)} = 1.3 \mu_v\) (``up''). Noise order flow \(u^{(n)}\) is i.i.d.\ Gaussian with distribution \(\mathcal{N}(0, \sigma_u^2)\) and \(\sigma_u = 50\). All learning agents are trained using PPO for 1000 episodes and evaluated on 30 holdout episodes. Each policy network consists of two hidden layers (64 units each) with Tanh activations. Prices are represented in cents to respect tick-size constraints, while rewards are computed in dollars to control numerical scale. We consider two global observation regimes: \texttt{OTC} mode (LOB hidden; only the transaction price observable) and \texttt{Exchange} mode (LOB observable).

\subsection{Equilibrium validation via price discovery}\label{exp:Kyle}

We validate equilibrium consistency by studying the extended Kyle benchmark, obtained by removing the liquidity trader so that the learning agents are the informed trader and the \(M\) market makers. The objective of this subsection is not to claim convergence to a Nash equilibrium, but rather to test whether the learned dynamics reproduce the structural signature of the benchmark equilibrium, namely, the gradual incorporation of private information into prices, under a policy class aligned with the theoretical linear parameterization (Eqs.~\ref{eq:IT-linear-param}--\ref{eq:MM-linear-param}). We also diagnose systematic failures when the policy class is overly expressive.

\subsubsection{Extended Kyle benchmark (linear parameterization)}

We measure price discovery using two standard diagnostics. Let \(v\) denote the fundamental value and define the pricing error \(e^{(n)} := v - \bar{p}^{(n)}\). First, we estimate an AR(1) model
\[
    e^{(n)} = \phi e^{(n-1)} + \epsilon^{(n)},
\]
and report the implied half-life
\[
    n_{\frac{1}{2}} = -\frac{\log 2}{\log |\phi|}.
\]
Second, we run a Kyle-style regression
\[
    \Delta \bar{p}^{(n)} := \bar{p}^{(n)} - \bar{p}^{(n-1)} = \lambda q^{(n)} + \xi^{(n)},
\]
where \(q^{(n)}\) denotes net order flow and \(\xi^{(n)}\) is the regression residual.

\begin{figure}[h!]
    \begin{center}
        \begin{minipage}{0.95\textwidth}
            \centering
            \subfigure[Opening price lower than the fundamental value.]{
            \resizebox*{0.45\textwidth}{!}{\includegraphics{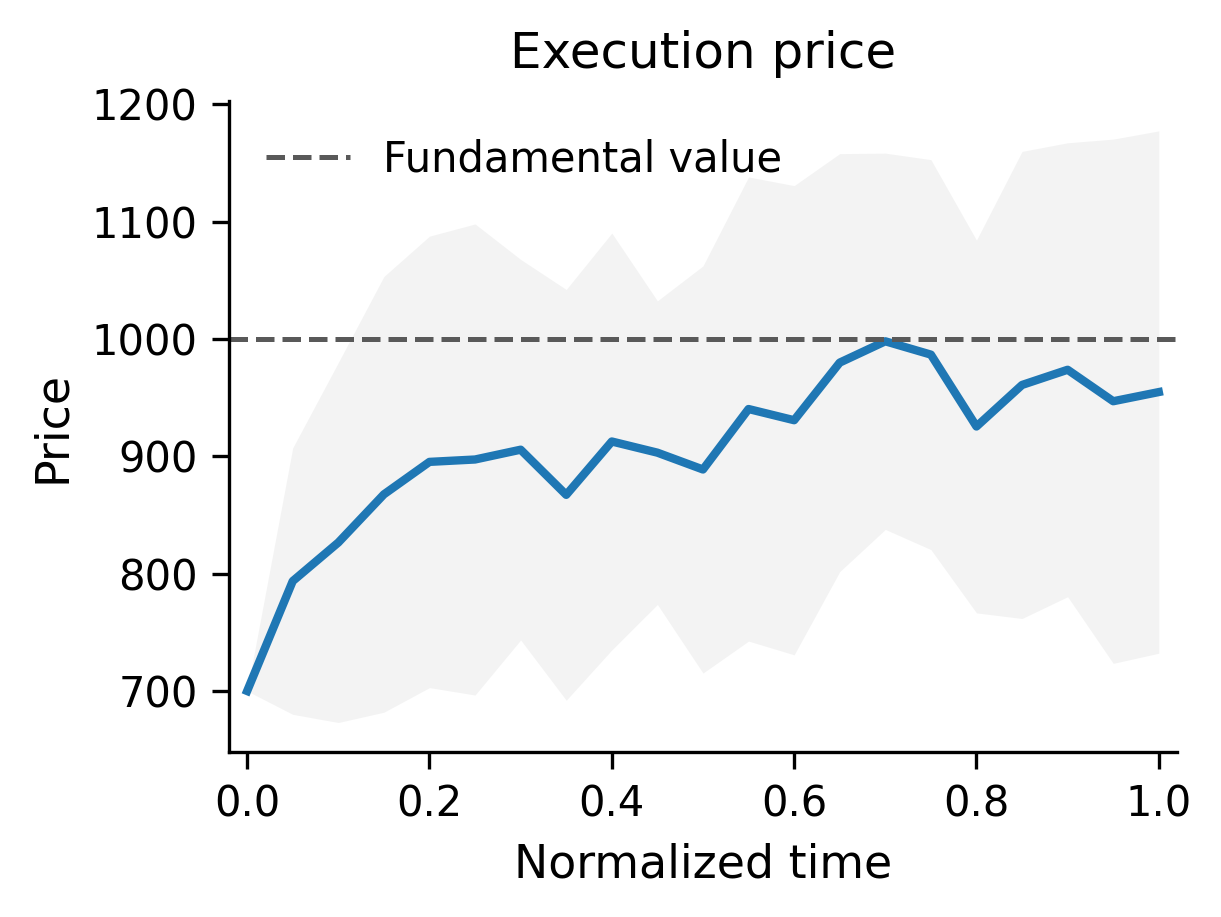}}\label{subfig:linear-20-MM-LOB-0-down}}
            \subfigure[Opening price higher than the fundamental value.]{
            \resizebox*{0.45\textwidth}{!}{\includegraphics{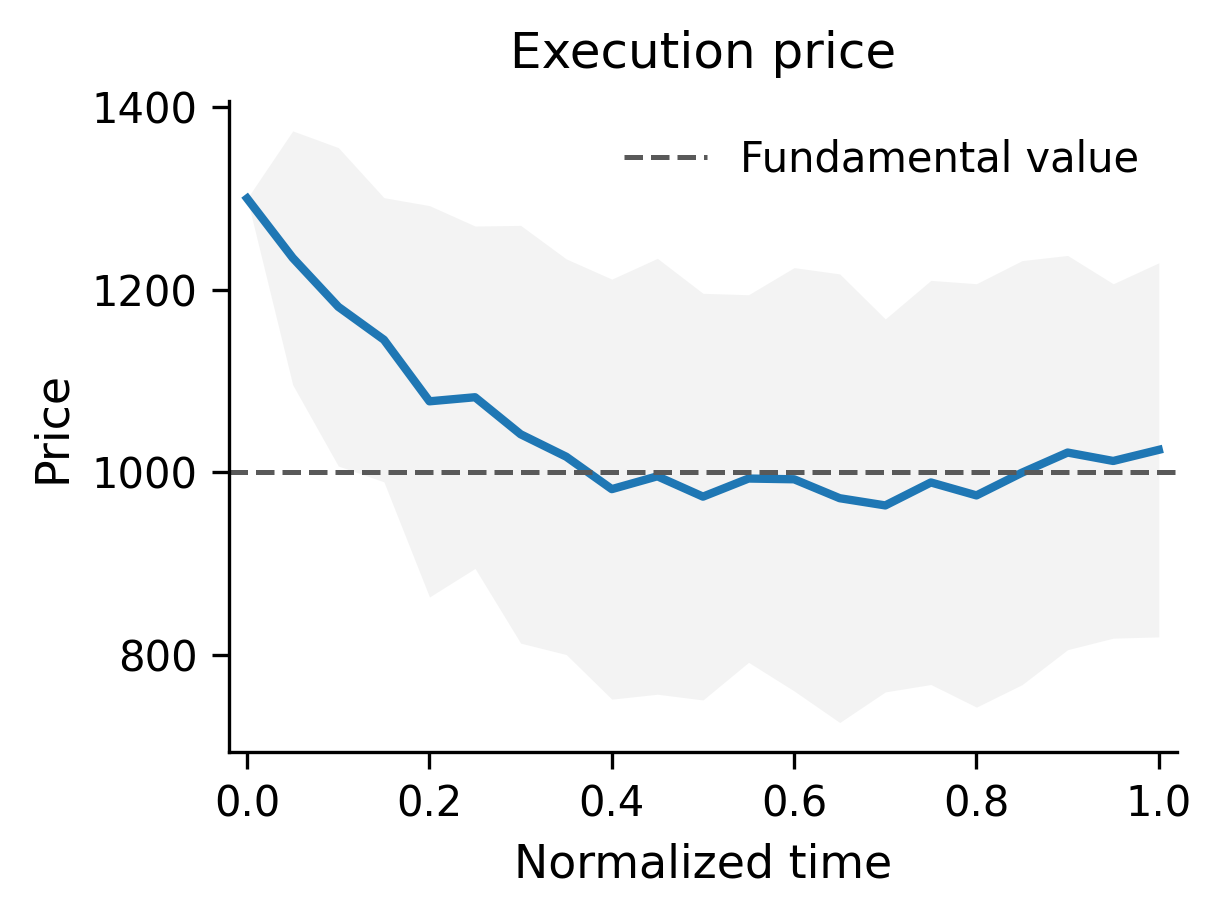}}\label{subfig:linear-20-MM-LOB-0-up}}
            \caption{Linear policy parameterization with 20 market makers and no LOB information revealed. 
            Each panel shows the evolution of transaction prices when the opening price is initialized below (a) or above (b) the fundamental value. The fundamental value is marked by the red dashed line. Across evaluation episodes, prices gradually converge toward the fundamental value, indicating that even with limited observability, market participants collectively recover informational efficiency. 
            \label{fig:linear-20-MM-LOB-0}}
        \end{minipage}
    \end{center}
\end{figure}

\begin{table}[h!]
\begin{center}
\begin{minipage}{\textwidth}
\setlength{\tabcolsep}{10pt}
\tbl{Price discovery experiment results under linear policy parameterization. Each configuration varies in the number of market makers (N MM), visibility of the LOB (0 hidden, 1 fully visible), and whether the initial price is below (down) or above (up) the fundamental value. Each configuration reports the estimated AR(1) coefficient (\(\phi\)), its p-value, the corresponding half-life, the Kyle regression coefficient (\(\lambda\)), and its p-value.}
{\begin{tabular}{@{}lccccccc@{}}
    \toprule
     \textbf{N MM} & \textbf{LOB} & \textbf{Mode} &
    \(\boldsymbol{\phi}\) & \textbf{p($\phi$)} & \textbf{Half-life} &
    \(\boldsymbol{\lambda}\) & \textbf{p($\lambda$)} \\
    \colrule
     20 & 0 & down & 0.745 & 3.18e-88 & 2.36 & 2.30 & 0.000 \\
     20 & 0 & up   & 0.804 & 0.000    & 3.18 & 2.78 & 7.05e-80 \\
     20 & 1 & down & 0.774 & 1.23e-249 & 2.70 & 2.88 & 0.000 \\
     20 & 1 & up   & 0.827 & 0.000    & 3.64 & 2.64 & 1.66e-41 \\
     2  & 0 & down & 0.232 & 7.09e-09 & 0.47 & 5.01 & 0.000 \\
     2  & 0 & up   & 0.416 & 2.36e-31 & 0.79 & 5.06 & 7.72e-124 \\
     2  & 1 & down & 0.503 & 3.94e-53 & 1.01 & 5.13 & 0.000 \\
     2  & 1 & up   & 0.625 & 1.15e-128 & 1.48 & 3.66 & 5.75e-85 \\
    \botrule
\end{tabular}}
\label{tab:price-discovery-linear}
\end{minipage}
\end{center}
\end{table}

\noindent Figure~\ref{fig:linear-20-MM-LOB-0} shows that under linear policy parameterization, transaction prices converge smoothly toward \(v\) under both initial conditions. Table~\ref{tab:price-discovery-linear} demonstrates the \textbf{robustness of the equilibrium structure} across different configurations. In particular, the estimated \(\phi\) is significantly less than one and the half-life is finite, consistent with progressive incorporation of private information into prices. The Kyle regression slope \(\lambda\) is positive and strongly significant, indicating that market makers systematically adjust quotes in the direction implied by order imbalance. This reproduces the qualitative hallmark of the multi-period Kyle equilibrium, i.e., gradual price discovery, under a theoretically aligned policy class. Across linear configurations, three consistent patterns emerge:
\begin{enumerate}
    \item \textbf{Fewer market makers accelerate convergence.} With smaller \(M\), the estimated half-life decreases, reflecting stronger per-agent price adjustment and reduced competitive dilution of impact.
    \item \textbf{LOB visibility affects speed but not direction.} Revealing LOB information changes the rate at which prices incorporate information, but the qualitative convergence toward \(v\) remains.
    \item \textbf{Symmetry in up/down initialization.} Whether the opening price starts above or below \(v\), convergence remains directionally correct and statistically significant.
\end{enumerate}
Together, these ablation results support the interpretation that the learning dynamics recover a stable equilibrium-like structure under an appropriate policy constraint.

\subsubsection{Policy class stress test (nonlinear policies)}

\begin{figure}[h!]
    \begin{center}
        \begin{minipage}{0.95\textwidth}
            \centering
            \subfigure[Opening price lower than the fundamental value.]{
            \resizebox*{0.45\textwidth}{!}{\includegraphics{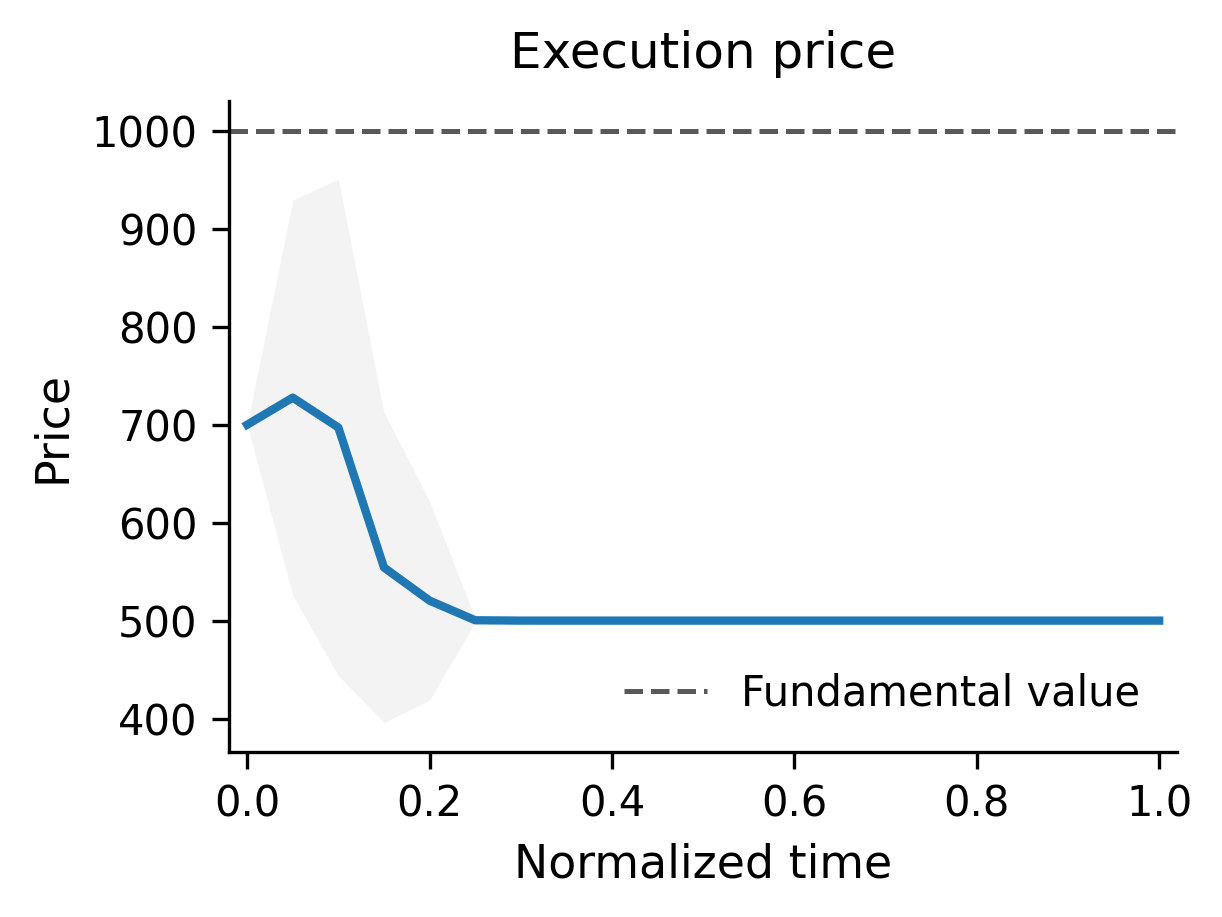}}\label{subfig:nonlinear-20-MM-LOB-0-down}}
            \subfigure[Opening price higher than the fundamental value.]{
            \resizebox*{0.45\textwidth}{!}{\includegraphics{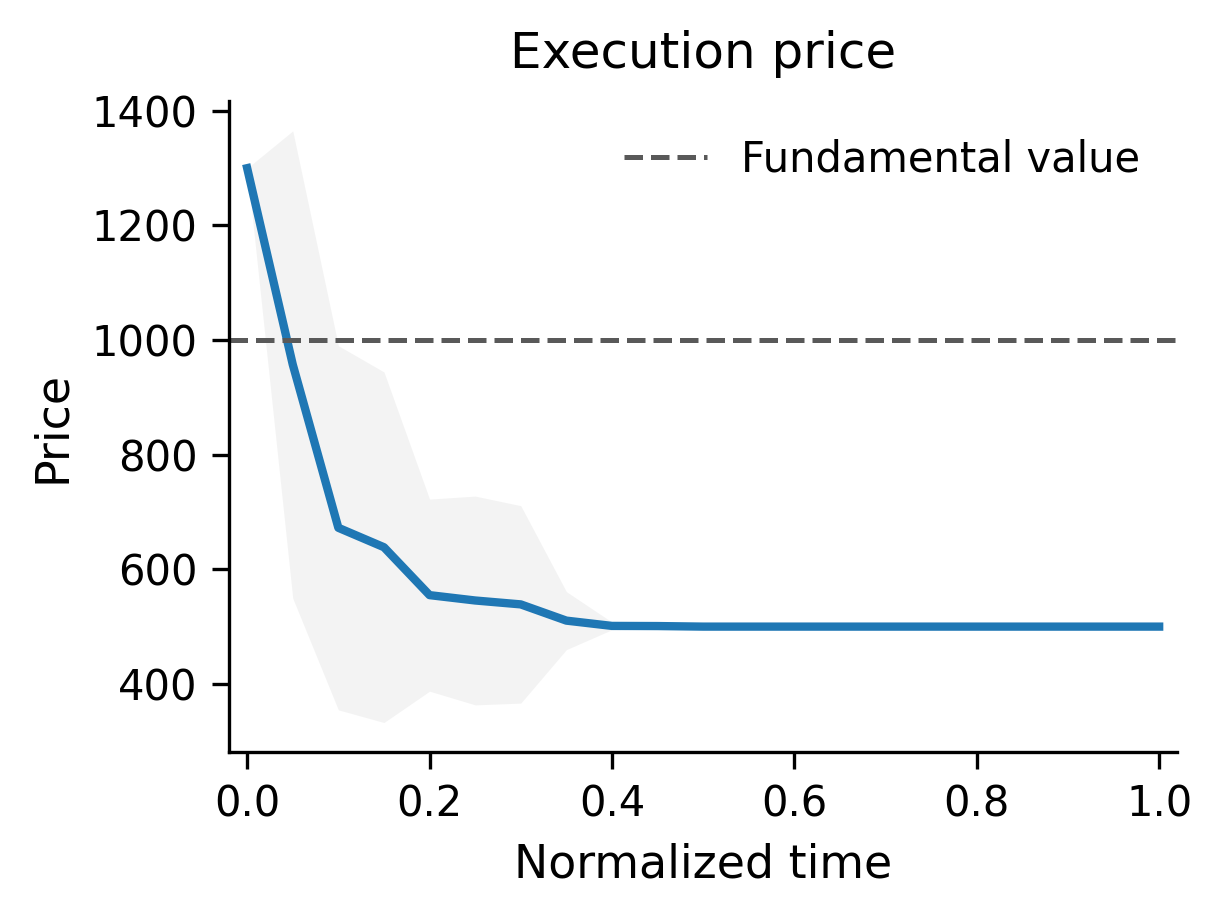}}\label{subfig:nonlinear-20-MM-LOB-0-up}}
            \caption{Nonlinear policy parameterization with 20 market makers and no LOB information revealed. 
            Panels (a) and (b) correspond to opening prices initialized below or above the fundamental value, respectively. The fundamental value is marked by the red dashed line. With the more flexible nonlinear policy network, the informed trader dominated the market dynamics, persistently driving transaction prices toward the lower boundary of the admissible price range regardless of the initial condition.
            \label{fig:nonlinear-20-MM-LOB-0}}
        \end{minipage}
    \end{center}
\end{figure}

\begin{table}[h!]
\begin{center}
\begin{minipage}{\textwidth}
\setlength{\tabcolsep}{11pt}
\tbl{Price discovery experiment results under nonlinear policy parameterization. Each configuration varies in the number of market makers (N MM), visibility of the LOB (0 hidden, 1 fully visible), and whether the initial price is below (down) or above (up) the fundamental value. Each configuration reports the estimated AR(1) coefficient (\(\phi\)), its p-value, the corresponding half-life, the Kyle regression coefficient (\(\lambda\)), and its p-value.}
{\begin{tabular}{@{}lccccccc@{}}
    \toprule
    \textbf{N MM} & \textbf{LOB} & \textbf{Mode} &
    \(\boldsymbol{\phi}\) & \textbf{p($\phi$)} & \textbf{Half-life} &
    \(\boldsymbol{\lambda}\) & \textbf{p($\lambda$)} \\
    \colrule
     20 & 0 & down & 0.995 & 0.000 & 145.41 & -0.14 & 0.139 \\
     20 & 0 & up   & 0.943 & 0.000 & 11.82  & 0.68 & 0.317 \\
     20 & 1 & down & 0.961 & 0.000 & 17.59  & 0.45 & 0.062 \\
     20 & 1 & up   & 1.009 & 0.000 & -81.16 & 0.15 & 0.071 \\
     2  & 0 & down & -0.058 & 0.193 & 0.24 & 3.25 & 0.000 \\
     2  & 0 & up   & 0.074 & 0.110 & 0.27 & 2.81 & 0.017 \\
     2  & 1 & down & 0.004 & 0.926 & 0.12 & 6.78 & 0.001 \\
     2  & 1 & up   & -0.051 & 0.289 & 0.23 & 14.89 & 1.98e-49 \\
    \botrule
\end{tabular}}
\label{tab:price-discovery-nonlinear}
\end{minipage}
\end{center}
\end{table}

When policies are nonlinear and unconstrained, the equilibrium structure breaks in a systematic way. Figure~\ref{fig:nonlinear-20-MM-LOB-0} shows that prices no longer converge reliably toward \(v\); instead, the informed trader can induce persistent deviations, often driving transaction prices toward boundary levels. In terms of diagnostics, Table~\ref{tab:price-discovery-nonlinear} shows that the AR(1) decay becomes weak or unstable (with \(\phi\) close to or exceeding one in several configurations), and the Kyle regression coefficient loses robustness and statistical significance.

\paragraph{Interpretation}

With an overly expressive policy class, agents can exploit predictability and boundary effects that are not present in the stylized analytical benchmark, leading to systematic failure modes rather than mere noisy instability. This experiment therefore delineates the regime in which the equilibrium proxy is credible (linear policy class) from the regime in which strategic exploitation dominates.

\subsection{Optimal execution under endogenous liquidity}

We now study execution in the full trading game with an informed trader, a liquidity trader, \(M\) market makers, and a representative noise trader. Unlike the exogenous-impact benchmark, where \(\{\lambda^{(n)}\}_{n = 1}^N\) is fixed, liquidity is now generated endogenously by the quoting policies of market makers. Consequently, the liquidity trader's execution cost becomes dependent on the joint policy profile.

\subsubsection{Experimental design and strategy set}

We evaluate five execution strategies for the liquidity trader, while the informed trader and market makers use the PPO policies obtained from multi-agent training.

\begin{enumerate}
    \item \textbf{PPO (joint MARL):} The liquidity trader uses the policy learned during joint multi-agent training.
    
    \item \textbf{VWAP (trajectory-based):} A fixed volume profile \(\{V^{(n)}\}_{n=1}^N\) is constructed from historical PPO evaluation trajectories (obtained from the PPO (joint MARL) holdout runs), and execution follows
    \[
        x_{\text{VWAP}}^{\text{LT}, (n)} = Q \cdot \frac{V^{(n)}}{\sum_{k=1}^N V^{(k)}}.
    \]

    \item \textbf{TWAP:} Inventory is executed evenly over time,
    \[
        x_{\text{TWAP}}^{\text{LT}, (n)} = \frac{Q}{N}.
    \]

    \item \textbf{Analytical (Kyle-style):} A Kyle-style time-varying impact path is computed from the linear recursive benchmark (Appendix~\ref{appendix:kyle-solution}) and substituted into the exogenous dynamic programming solution (Theorem~\ref{thm:liquidity-opt}) to obtain an execution schedule. This serves as a \emph{fixed-impact} benchmark and is not equilibrium-consistent under endogenous liquidity.

    \item \textbf{PPO-Single:} The informed trader and market makers are frozen at their learned policies, and only the liquidity trader is retrained using PPO (500 episodes), yielding a best-response style policy against fixed opponents.
\end{enumerate}
Performance is evaluated using \textbf{implementation shortfall (IS)}:
\[
    \text{IS} := \frac{1}{p^{(0)}}\left(\frac{\sum_{n=1}^N p^{(n)} x^{\text{LT}, (n)}}{Q} - p^{(0)}\right).
\]
Lower IS indicates better execution performance for an acquisition task. Unless stated otherwise, we set \(Q = 1000\), the terminal penalty \(\gamma = 10\), and consider \(\phi \in \{0, 0.01, 0.1\}\). We set \(\alpha = 0\) to align the permanent component with the Kyle-style benchmark and to isolate endogenous liquidity effects from mean-reversion effects.

\subsubsection{Linear policy parameterization (constrained opponents)}

\begin{table}
\begin{center}
\begin{minipage}{\textwidth}
\setlength{\tabcolsep}{5pt}
\tbl{Implementation shortfall (IS) comparison under linear policy parameterization. 
Values denote mean \(\pm\) standard deviation over 30 evaluation episodes, normalized to the opening price \(p^{(0)}\). 
Lower values indicate better performance. ``N/A'' in the PPO column denotes cases where the agent did not execute any acquisition during the trading horizon.}
{\begin{tabular}{@{}lccccccc@{}}
\toprule
\textbf{LOB} & \(\boldsymbol{\phi}\) & \textbf{Mode} &
\textbf{PPO} & \textbf{VWAP} & \textbf{TWAP} & \textbf{Analytical} & \textbf{PPO-Single} \\
\colrule
0 & 0 & down & N/A & \(0.698 \pm 0.046\) & \(0.627 \pm 0.046\) & \(0.807 \pm 0.141\) & \(\mathbf{0.597 \pm 0.046}\) \\
0 & 0 & up & N/A & \(0.008 \pm 0.035\) & \(\mathbf{-0.065 \pm 0.026}\) & \(0.108 \pm 0.020\) & \(-0.014 \pm 0.026\) \\
0 & 0.01 & down & \(0.876 \pm 0.012\) & \(0.834 \pm 0.021\) & \(\mathbf{0.750 \pm 0.073}\) & \(0.775 \pm 0.084\) & \(0.801 \pm 0.079\) \\
0 & 0.01 & up & \(0.149 \pm 0.003\) & \(0.082 \pm 0.012\) & \(\mathbf{-0.028 \pm 0.038}\) & \(0.117 \pm 0.015\) & \(0.055 \pm 0.032\) \\
0 & 0.1 & down & \(0.841 \pm 0.045\) & \(0.811 \pm 0.027\) & \(0.714 \pm 0.066\) & \(0.951 \pm 0.037\) & \(\mathbf{0.703 \pm 0.022}\) \\
0 & 0.1 & up & \(0.113 \pm 0.008\) & \(0.072 \pm 0.016\) & \(\mathbf{-0.007 \pm 0.038}\) & \(0.134 \pm 0.008\) & \(0.105 \pm 0.007\) \\
1 & 0 & down & \(1.014 \pm 0.053\) & \(0.747 \pm 0.059\) & \(\mathbf{0.740 \pm 0.088}\) & \(0.784 \pm 0.103\) & \(0.795 \pm 0.130\) \\
1 & 0 & up & \(0.153 \pm 0.001\) & \(0.112 \pm 0.054\) & \(\mathbf{0.079 \pm 0.059}\) & \(0.107 \pm 0.016\) & \(0.111 \pm 0.037\) \\
1 & 0.01 & down & \(0.936 \pm 0.060\) & \(0.775 \pm 0.088\) & \(\mathbf{0.703 \pm 0.100}\) & \(0.790 \pm 0.099\) & \(0.799 \pm 0.009\) \\
1 & 0.01 & up & \(0.087 \pm 0.028\) & \(\mathbf{0.046 \pm 0.035}\) & \(0.050 \pm 0.030\) & \(0.118 \pm 0.020\) & \(0.144 \pm 0.001\) \\
1 & 0.1 & down & \(0.886 \pm 0.013\) & \(0.767 \pm 0.021\) & \(\mathbf{0.500 \pm 0.019}\) & \(0.923 \pm 0.055\) & \(0.548 \pm 0.028\) \\
1 & 0.1 & up & \(0.154 \pm 0.000\) & \(-0.172 \pm 0.009\) & \(\mathbf{-0.178 \pm 0.010}\) & \(0.135 \pm 0.008\) & \(-0.052 \pm 0.023\) \\
\botrule
\end{tabular}}
\label{tab:imp-shortfall-linear}
\end{minipage}
\end{center}
\end{table}

Table~\ref{tab:imp-shortfall-linear} reports the results under linear policy parameterization for the informed trader and the \(M\) market makers.\\ \\  Three findings are robust across configurations:

\begin{enumerate}
    \item \textbf{TWAP is the most robust baseline.} A non-reactive schedule performs well when opponents are constrained.
    
    \item \textbf{The analytical schedule under-performs.} Although optimal under exogenous fixed impact, it becomes systematically inferior once liquidity is endogenous. This provides a direct indication of structural misspecification: the assumed impact path does not match the impact induced by strategic quoting.
    
    \item \textbf{Joint MARL PPO is exploited.} The liquidity trader's jointly trained policy performs poorly in several configurations, indicating that during training, other agents adapt to the liquidity trader's behavior and learn responses that are unfavorable to her objective.
\end{enumerate}

\paragraph{Interpretation}

Under constrained policy classes, the informed trader's influence on market dynamics is limited, and execution is best served by stable schedules that reduce predictability and avoid triggering strong opponent adaptation. The failure of the analytical schedule already indicates that ``plugging in'' fixed-impact solutions is unreliable in endogenous markets.

\subsubsection{Nonlinear policy parameterization (expressive opponents)}

\begin{figure}[h!]
    \begin{center}
        \begin{minipage}{\linewidth}
            \centering
            \subfigure[PPO]{
            \resizebox*{0.48\textwidth}{!}{\includegraphics{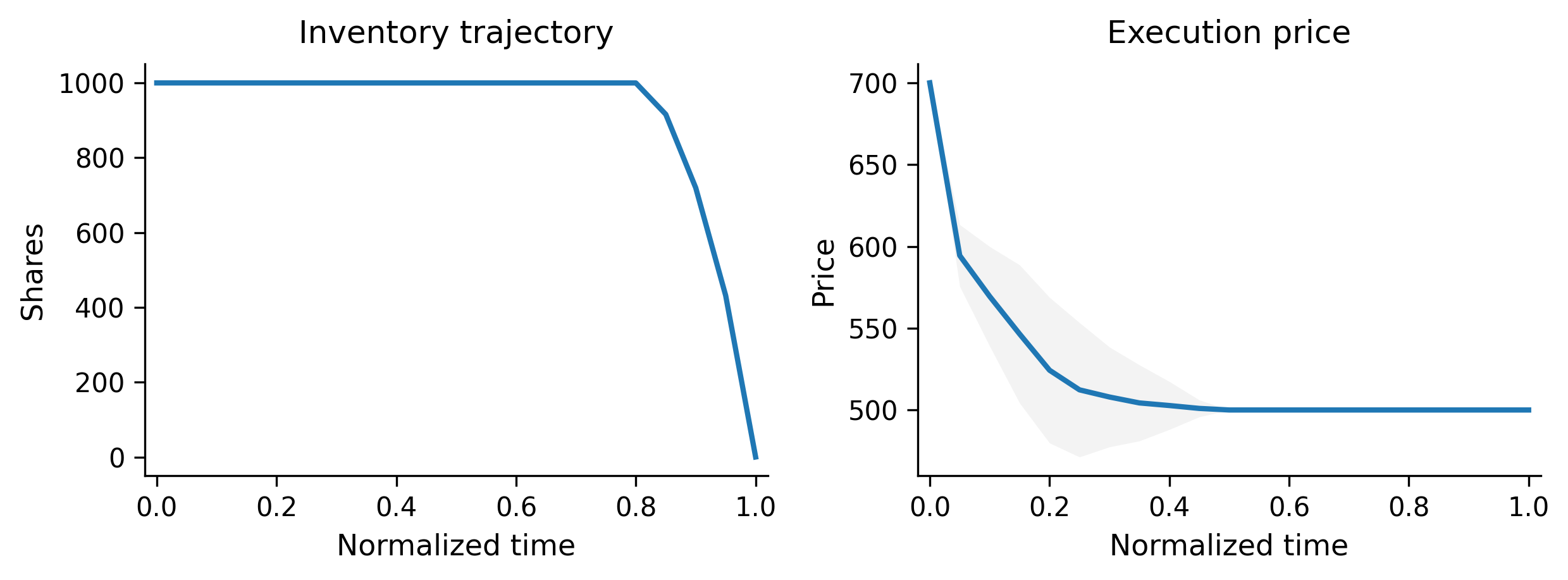}}\label{subfig:nonlinear-down-ppo}}
            \subfigure[VWAP]{
            \resizebox*{0.48\textwidth}{!}{\includegraphics{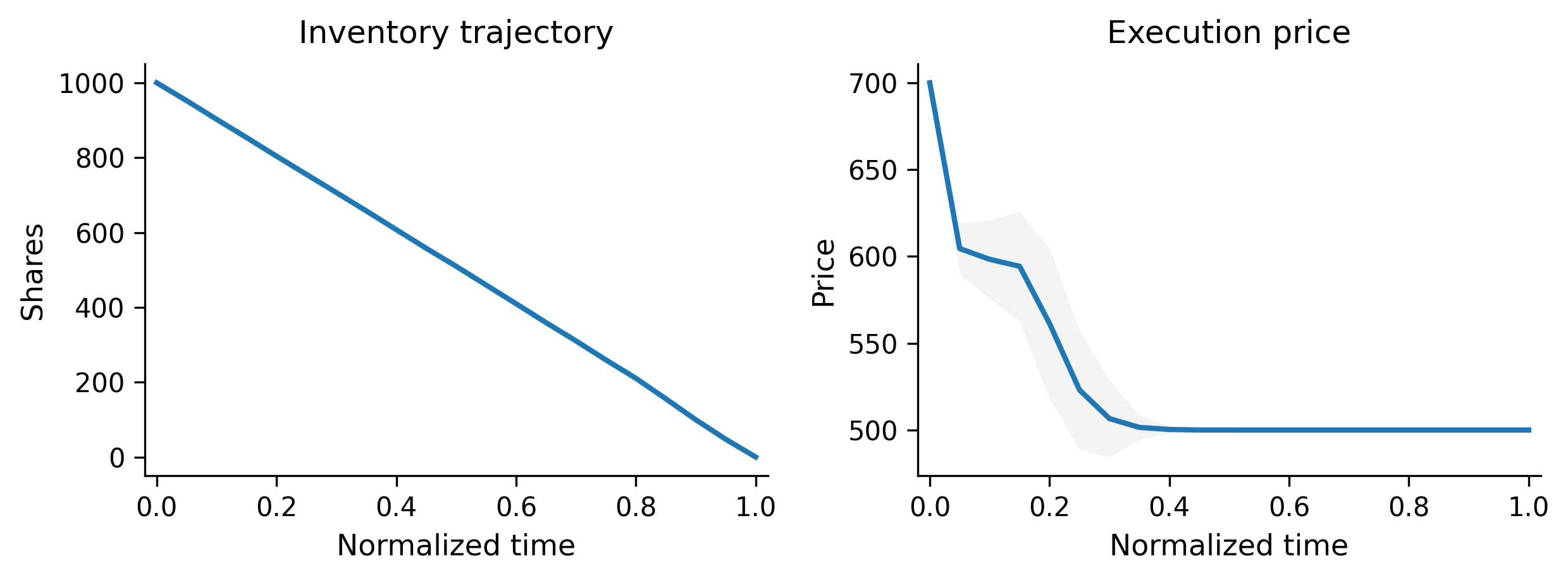}}\label{subfig:nonlinear-down-vwap}}
            \subfigure[TWAP]{
            \resizebox*{0.48\textwidth}{!}{\includegraphics{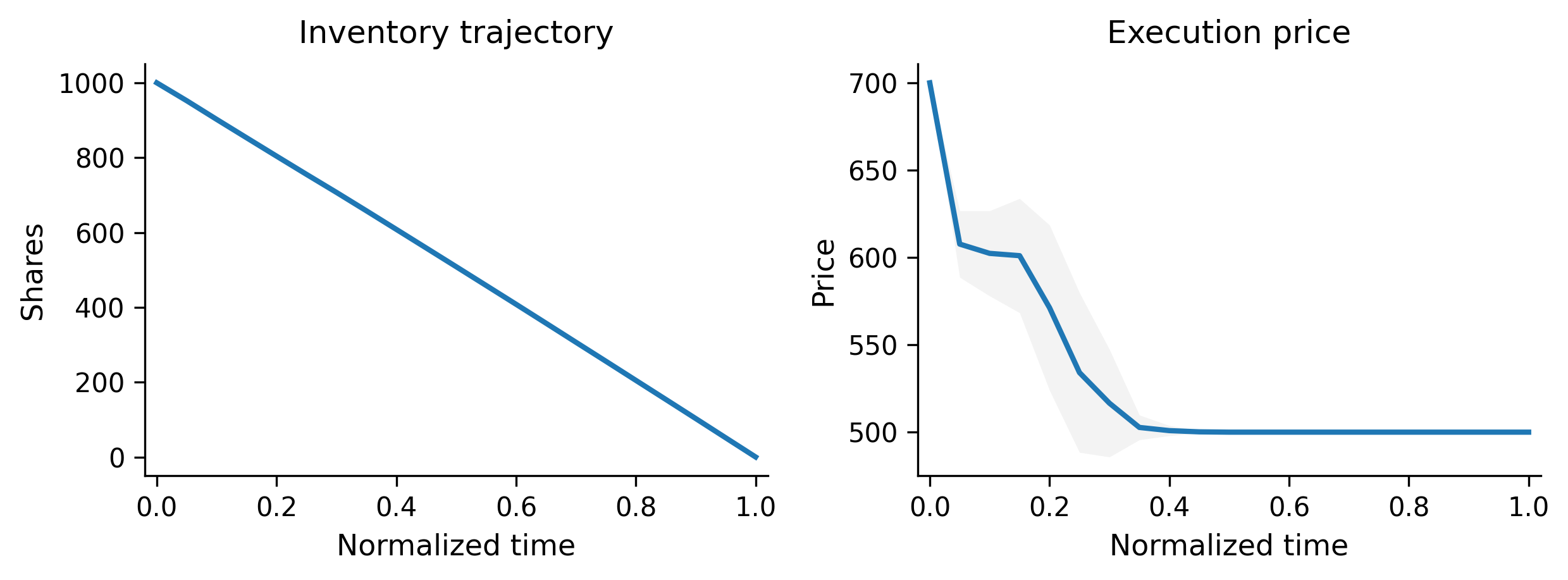}}\label{subfig:nonlinear-down-twap}}
            \subfigure[Analytical]{
            \resizebox*{0.48\textwidth}{!}{\includegraphics{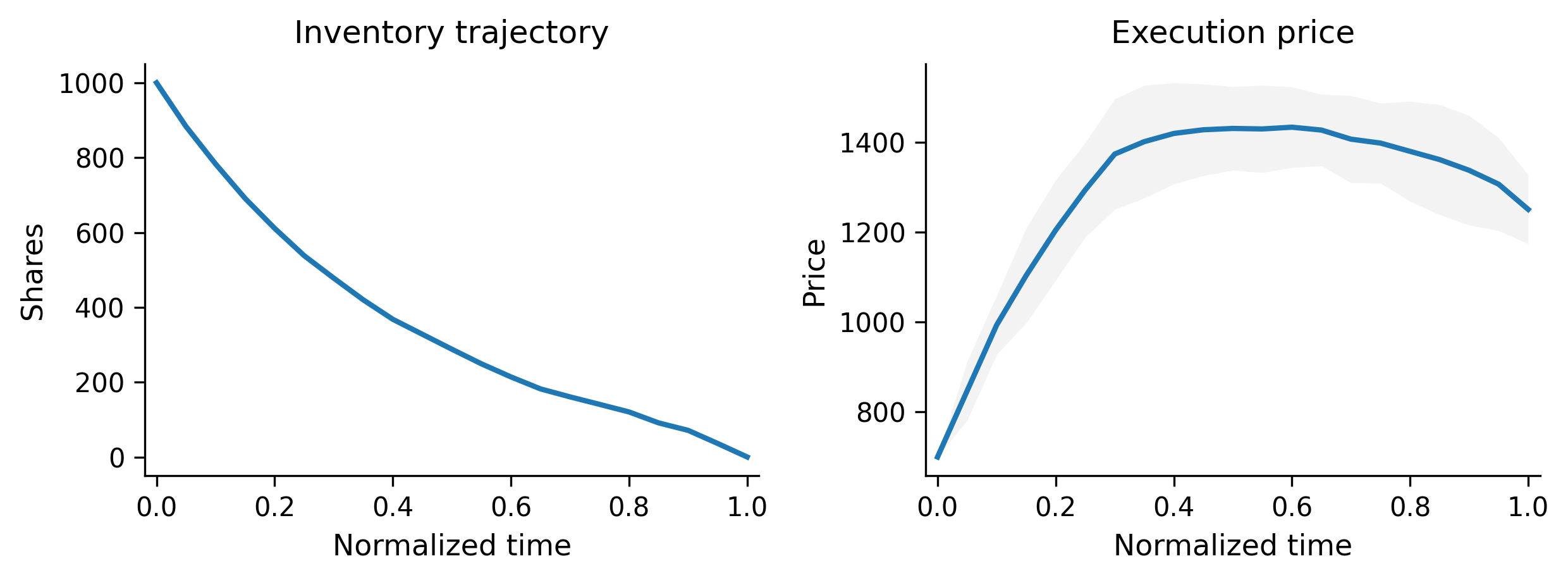}}\label{subfig:nonlinear-down-analytical}}
            \subfigure[PPO Single]{
            \resizebox*{0.48\textwidth}{!}{\includegraphics{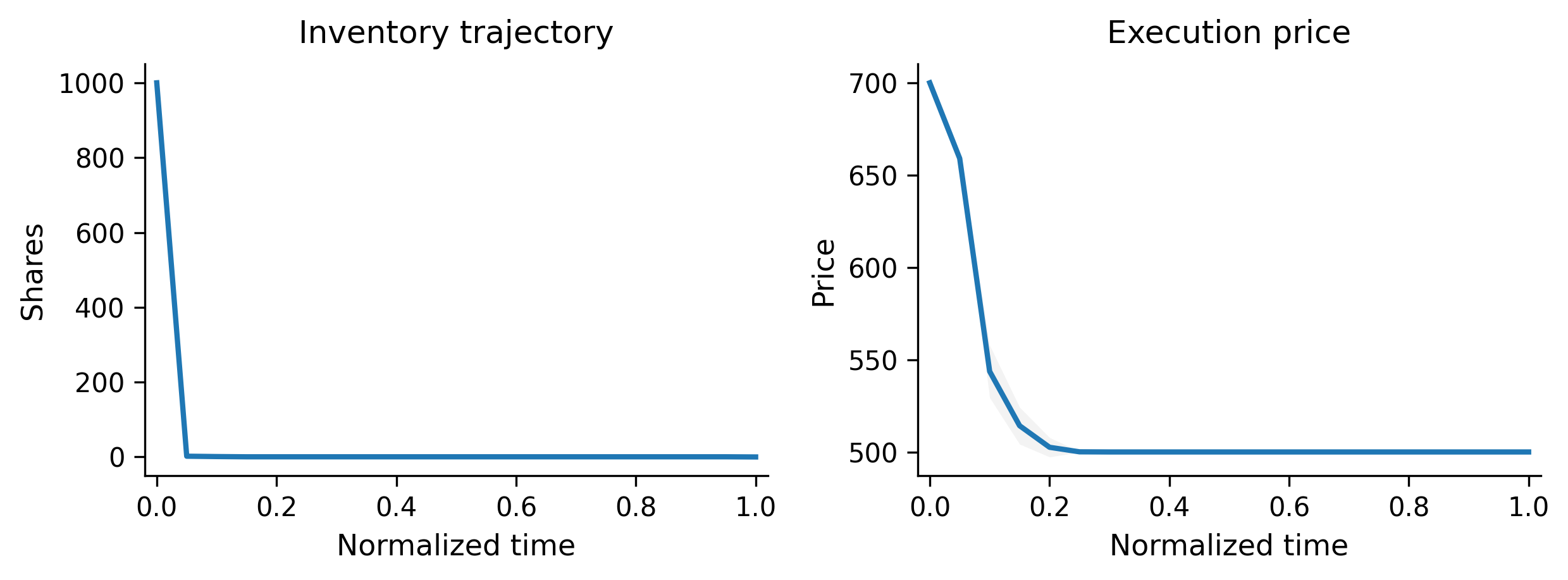}}\label{subfig:nonlinear-down-ppo-single}}
            \caption{
            Nonlinear policy parameterization with 20 market makers and no LOB information revealed. 
            The liquidity trader’s risk aversion is set to \(\phi = 0.01\), and the opening price is initialized below the fundamental value. 
            Each panel reports the price trajectory (left) and unfilled inventory process (right) averaged over 30 evaluation episodes. 
            Among the five strategies, the PPO policy achieves the lowest implementation shortfall by waiting for prices to reach the floor level before initiating acquisition. 
            In contrast, the Analytical strategy performs the worst, as it is the only one that induces an upward price movement during execution.
            \label{fig:nonlinear-down-strategies}}
        \end{minipage}
    \end{center}
\end{figure}

\begin{figure}[h!]
    \begin{center}
        \begin{minipage}{\linewidth}
            \centering
            \subfigure[PPO]{
            \resizebox*{0.48\textwidth}{!}{\includegraphics{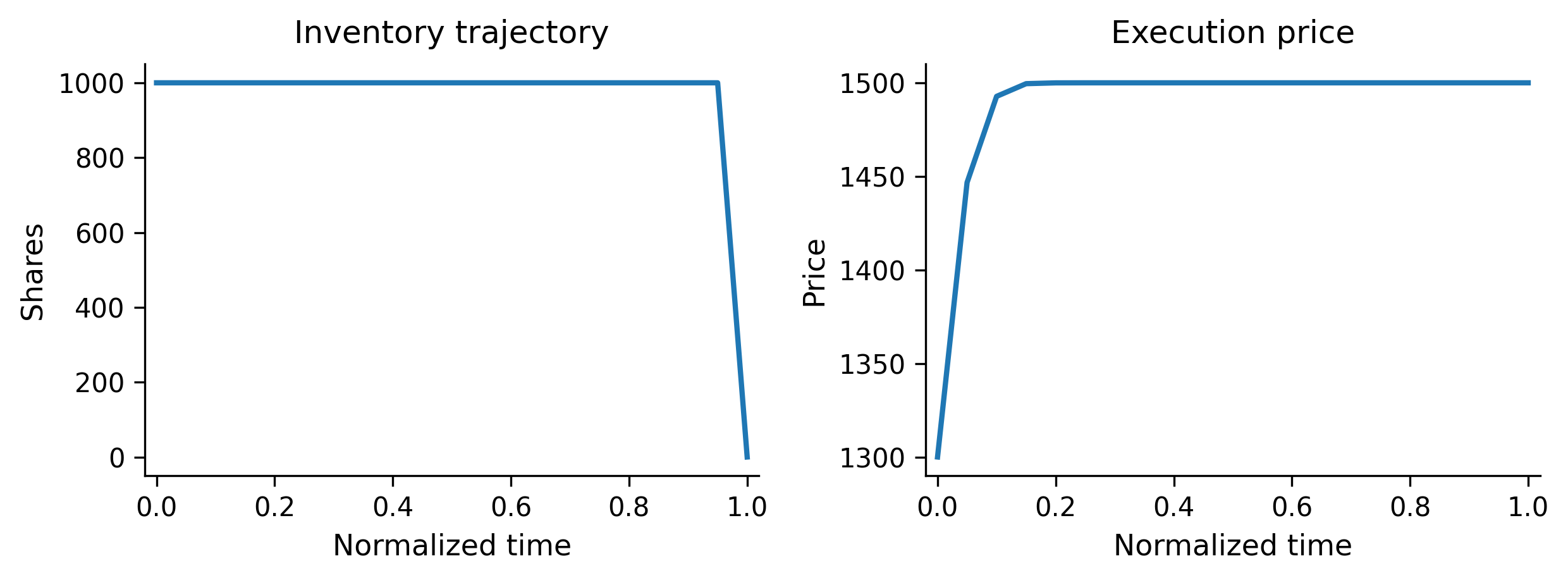}}\label{subfig:nonlinear-up-ppo}}
            \subfigure[VWAP]{
            \resizebox*{0.48\textwidth}{!}{\includegraphics{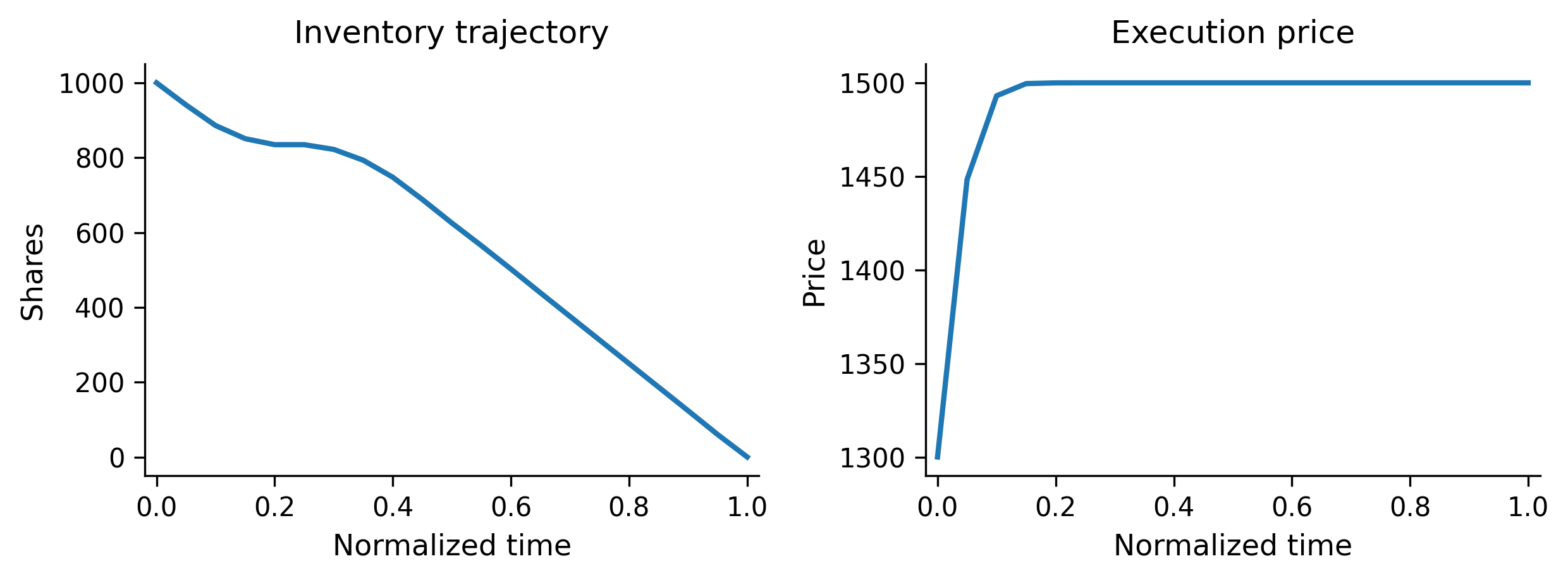}}\label{subfig:nonlinear-up-vwap}}
            \subfigure[TWAP]{
            \resizebox*{0.48\textwidth}{!}{\includegraphics{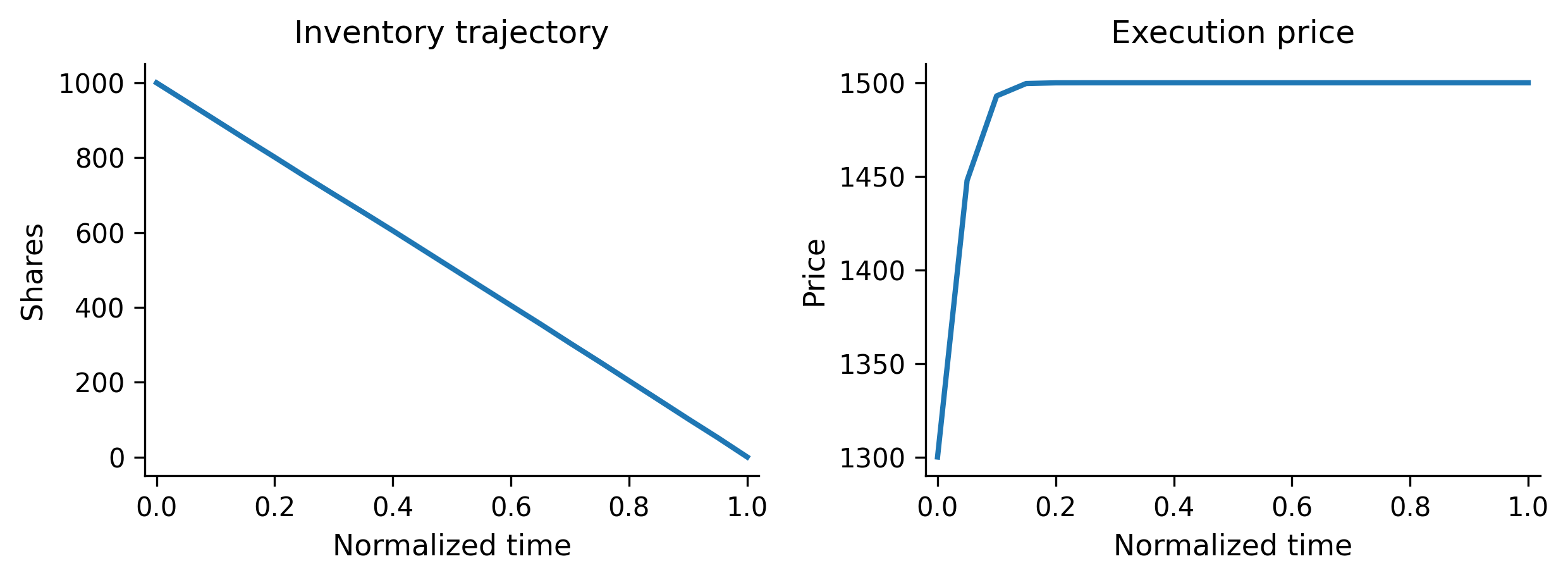}}\label{subfig:nonlinear-up-twap}}
            \subfigure[Analytical]{
            \resizebox*{0.48\textwidth}{!}{\includegraphics{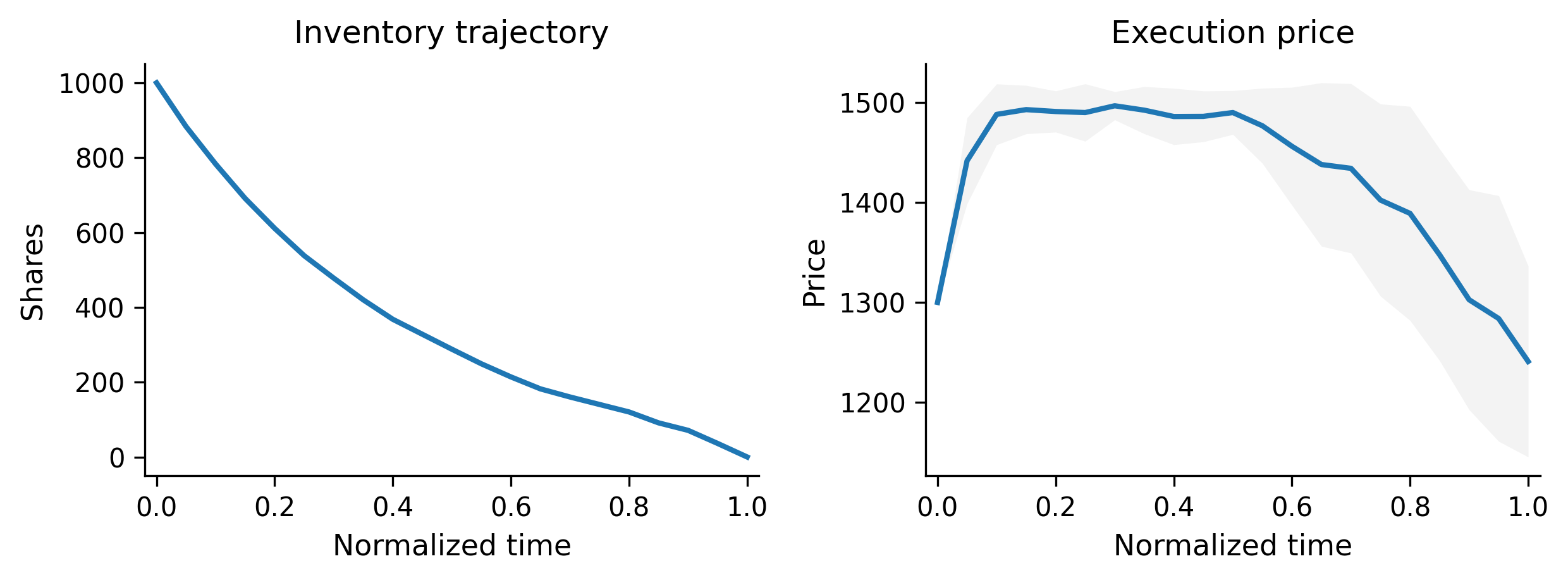}}\label{subfig:nonlinear-up-analytical}}
            \subfigure[PPO Single]{
            \resizebox*{0.48\textwidth}{!}{\includegraphics{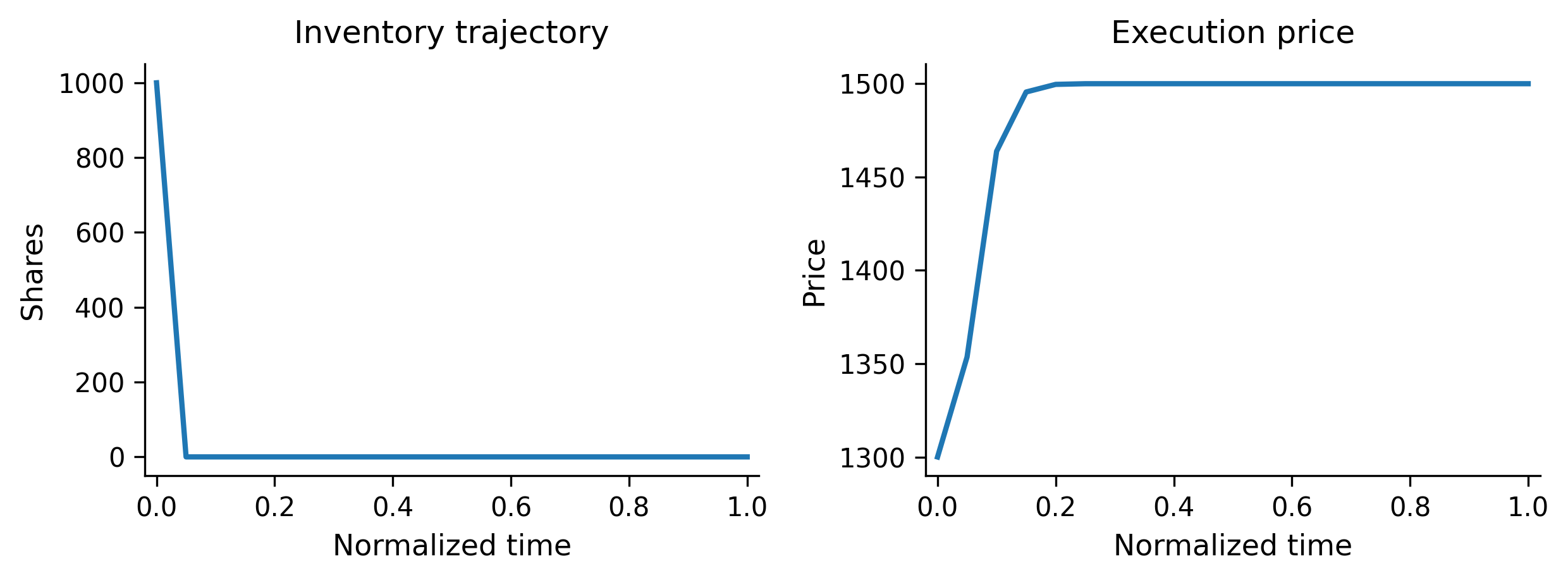}}\label{subfig:nonlinear-up-ppo-single}}
            \caption{
            Nonlinear policy parameterization with 20 market makers and no LOB information revealed. 
            The liquidity trader’s risk aversion is set to \(\phi = 0.01\), and the opening price is initialized above the fundamental value. 
            Each panel reports the price trajectory (left) and unfilled inventory process (right), averaged over 30 evaluation episodes. 
            Among the five strategies, the PPO-Single policy achieves the lowest implementation shortfall, despite not waiting for prices to decline before trading. 
            The Analytical strategy, by inducing downward price adjustments during execution, performs notably better in this setting, ranking as the second-best approach.
            \label{fig:nonlinear-up-strategies}}
        \end{minipage}
    \end{center}
\end{figure}

Compared with the linear case, Table~\ref{tab:imp-shortfall-nonlinear} and Figures~\ref{fig:nonlinear-down-strategies}--\ref{fig:nonlinear-up-strategies} show that nonlinear parameterization amplifies endogenous effects:

\begin{enumerate}
    \item \textbf{PPO-Single is superior.} Post-training against fixed opponents significantly improves the strategy, yielding lower IS. This indicates that once opponents stop adapting, the liquidity trader can exploit the induced market dynamics more effectively.
    
    \item \textbf{Analytical remains unreliable.} The fixed-impact schedule does not consistently perform well, even when it occasionally benefits from favorable price drift; its performance remains unstable across initial conditions.
    
    \item \textbf{TWAP loses robustness.} With expressive opponents, non-reactive schedules are no longer robust to strategic adaptation.
    
    \item \textbf{Market makers intensify adaptation.} The dispersion of outcomes and sensitivity to initial conditions both increase, consistent with endogenous liquidity being shaped by strategic responses rather than by a fixed coefficient process.
\end{enumerate}
\vspace{-2mm}

\begin{table}[h!]
\begin{center}
\begin{minipage}{\textwidth}
\setlength{\tabcolsep}{5pt}
\tbl{Implementation shortfall (IS) comparison under non-linear policy parameterization. 
Values denote mean \(\pm\) standard deviation over 30 evaluation episodes, normalized to the opening price \(p^{(0)}\). 
Lower values indicate better performance.}
{\begin{tabular}{@{}lcccccccc@{}}
\toprule
\textbf{LOB} & \(\boldsymbol{\phi}\) & \textbf{Mode} &
\textbf{PPO} & \textbf{VWAP} & \textbf{TWAP} & \textbf{Analytical} & \textbf{PPO-Single} \\
\colrule
0 & 0 & down & \(-0.236 \pm 0.061\) & \(-0.178 \pm 0.020\) & \(-0.182 \pm 0.024\) & \(0.805 \pm 0.075\) & \(\mathbf{-0.248 \pm 0.056}\) \\
0 & 0 & up & \(\mathbf{-0.152 \pm 0.007}\) & \(-0.142 \pm 0.005\) & \(-0.141 \pm 0.004\) & \(0.103 \pm 0.029\) & \(-0.147 \pm 0.006\) \\
0 & 0.01 & down & \(\mathbf{-0.286 \pm 0.000}\) & \(-0.258 \pm 0.010\) & \(-0.255 \pm 0.013\) & \(0.778 \pm 0.089\) & \(-0.061 \pm 0.003\) \\
0 & 0.01 & up & \(0.154 \pm 0.001\) & \(0.151 \pm 0.000\) & \(0.152 \pm 0.000\) & \(0.116 \pm 0.017\) & \(\mathbf{0.041 \pm 0.001}\) \\
0 & 0.1 & down & \(0.000 \pm 0.010\) & \(-0.271 \pm 0.009\) & \(\mathbf{-0.282 \pm 0.004}\) & \(0.916 \pm 0.065\) & \(-0.200 \pm 0.003\) \\
0 & 0.1 & up & \(0.051 \pm 0.006\) & \(-0.413 \pm 0.023\) & \(\mathbf{-0.432 \pm 0.018}\) & \(0.135 \pm 0.008\) & \(-0.144 \pm 0.010\) \\
1 & 0 & down & \(-0.278 \pm 0.003\) & \(-0.278 \pm 0.002\) & \(-0.278 \pm 0.003\) & \(0.783 \pm 0.091\) & \(\mathbf{-0.286 \pm 0.000}\) \\
1 & 0 & up & \(-0.348 \pm 0.121\) & \(-0.415 \pm 0.072\) & \(-0.387 \pm 0.082\) & \(0.116 \pm 0.012\) & \(\mathbf{-0.603 \pm 0.068}\) \\
1 & 0.01 & down & \(-0.229 \pm 0.016\) & \(-0.262 \pm 0.008\) & \(-0.265 \pm 0.003\) & \(0.824 \pm 0.076\) & \(\mathbf{-0.266 \pm 0.002}\) \\
1 & 0.01 & up & \(0.144 \pm 0.003\) & \(0.143 \pm 0.001\) & \(0.144 \pm 0.001\) & \(\mathbf{0.116 \pm 0.017}\) & \(0.150 \pm 0.000\) \\
1 & 0.1 & down & \(-0.217 \pm 0.025\) & \(-0.280 \pm 0.002\) & \(\mathbf{-0.281 \pm 0.001}\) & \(0.925 \pm 0.043\) & \(-0.168 \pm 0.006\) \\
1 & 0.1 & up & \(-0.052 \pm 0.004\) & \(-0.165 \pm 0.023\) & \(\mathbf{-0.170 \pm 0.009}\) & \(0.134 \pm 0.009\) & \(0.004 \pm 0.007\) \\
\botrule
\end{tabular}}
\label{tab:imp-shortfall-nonlinear}
\end{minipage}
\end{center}
\end{table}

\paragraph{Interpretation}

When agents have expressive policy classes, the execution problem becomes strongly policy-profile dependent. The notion of ``optimal execution under a fixed impact model'' therefore collapses, because the effective liquidity process is itself an equilibrium outcome.

\subsection{Degenerate case: no informed trader}\label{exp:LT-vs-MMs}

We finally remove the informed trader and consider the game between the liquidity trader and the market makers only. This experiment isolates the role of informational balance.
\begin{figure}[h!]
    \begin{center}
        \begin{minipage}{\linewidth}
            \centering
            \subfigure[Opening price lower than the fundamental value]{
            \resizebox*{0.48\textwidth}{!}{\includegraphics{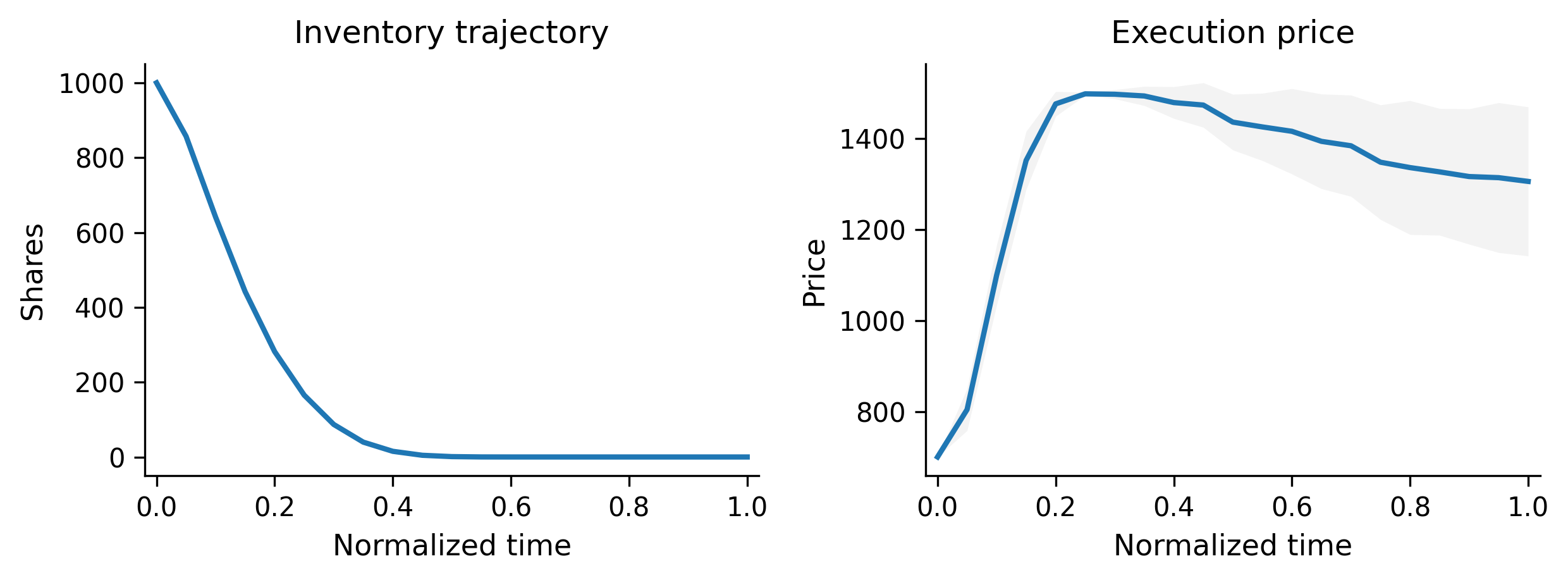}}\label{subfig:linear-down-LT-vs-MMs}}
            \subfigure[Opening price higher than the fundamental value]{
            \resizebox*{0.48\textwidth}{!}{\includegraphics{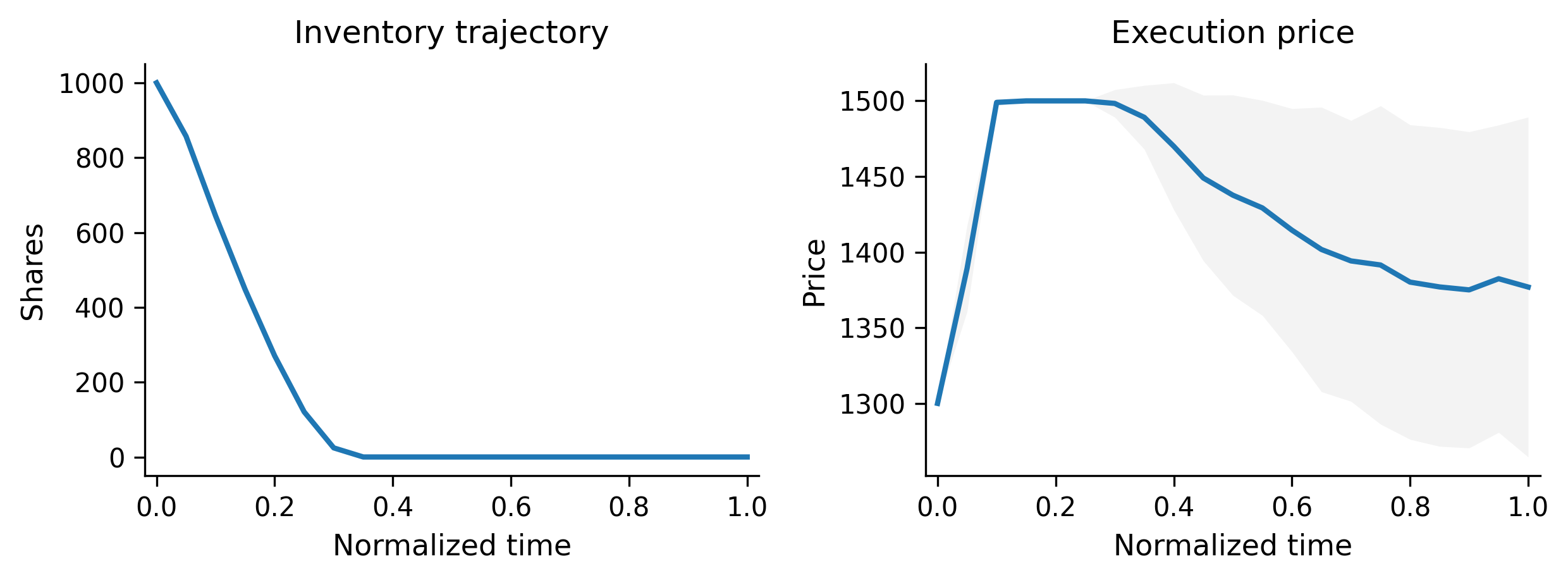}}\label{subfig:linear-up-LT-vs-MMs}}
            \caption{
                Price and unfilled-inventory dynamics for the liquidity trader trading against 20 market makers under linear policy parameterization, with no LOB information revealed. 
                The liquidity trader’s risk aversion is set to \(\phi = 0.01\). 
                Panels~(a) and~(b) correspond to opening prices initialized below and above the fundamental value, respectively. 
                In both cases, market makers rapidly adjust their quotes upward, limiting the trader’s ability to reduce transaction costs. 
                This behavior reflects the market makers’ adaptation to the trader’s predictable acquisition strategy, resulting in consistently adverse price movements.
                \label{fig:linear-LT-vs-MMs}
            }
        \end{minipage}
    \end{center}
\end{figure}
Figure~\ref{fig:linear-LT-vs-MMs} shows that in the absence of an informed trader, market makers anticipate a predictable forced acquisition. The resulting dynamics become degenerate: market makers rapidly move quotes against the liquidity trader, leaving the trader with limited ability to reduce execution costs through strategic adaptation. This degeneracy highlights a key distinction between exogenous and endogenous liquidity frameworks. Under exogenous liquidity, informational balance is largely irrelevant because price impact is fixed. Under endogenous liquidity, however, informational asymmetry and strategic anticipation directly determine the effective impact process. Without a countervailing informational force, the execution problem can collapse into a structurally unfavorable regime.

\section{Conclusion}\label{sec:discussion}

 In standard optimal execution models, liquidity and price impact are treated as exogenous, which is analytically convenient but inadequate when liquidity provision is strategic. In such settings, execution is no longer a single-agent control problem against a passive environment, but a stochastic game in which the trader’s actions affect and are affected by the evolving behavior of market makers. This paper develops an equilibrium-consistent framework for studying optimal execution under endogenous liquidity by combining a synchronized multi-agent limit-order-book simulator with a stochastic-game formulation and a MARL-based equilibrium approximation. Our main finding is that, under endogenous liquidity, market impact is policy-profile dependent: execution strategies influence the distribution of order flow faced by market makers; market makers adapt their quotes in response, and the resulting effective impact feeds back into execution costs. As a result, optimal strategies differ structurally from those implied by fixed-impact models, making equilibrium consistency essential for meaningful execution analysis. Although we do not claim convergence to an exact Nash equilibrium, the framework provides a practical and economically grounded approach to approximating strategic interaction in realistic LOB markets, and offers a foundation for future work on richer agent populations, alternative learning dynamics, and broader applications in agent-based financial AI.

\section*{Supplemental material}
\renewcommand{\thesubsection}{\Alph{subsection}}

\subsection{Proof of Theorem~\ref{thm:liquidity-opt}}

\begin{proof}
We proceed by backward induction.

\paragraph{Base Case (\(n = N\))} 
At the final period, we evaluate
\begin{align*}
    L^{(N)}(\tilde{p}^{(N-1)}, Q^{(N-1)}, Q^{(N)}) 
    = 
    \mathbb{E}_N \left[ p^{(N)} x^{\text{LT},(N)} + \phi \big(Q^{(N)}\big)^2 \right].
\end{align*}
To express this in terms of known quantities \((\tilde{p}^{(N-1)}, Q^{(N-1)}, Q^{(N)})\), substitute for \(p^{(N)}\) using 
\eqref{eq:methods-liquidity-permanent-price-impact} and 
\eqref{eq:methods-liquidity-temp-price-impact}:
\begin{align*}
    p^{(N)} 
    &= 
    \hat{p}^{(N)} + \lambda^{(N)} \big(x^{\text{LT},(N)} + u^{(N)}\big) \\
    &= 
    \alpha \hat{p}^{(N-1)} + (1 - \alpha) p^{(N-1)} + \epsilon^{(N)} 
    + 
    \lambda^{(N)} \big(x^{\text{LT},(N)} + u^{(N)}\big) \\
    &= 
    \alpha (\hat{p}^{(N-1)} - p^{(N-1)}) 
    + 
    \big(p^{(N-1)} + \epsilon^{(N)}\big) 
    + 
    \lambda^{(N)} \big(x^{\text{LT},(N)} + u^{(N)}\big) \\
    &= 
    -\alpha \lambda^{(N-1)} \big(x^{\text{LT},(N-1)} + u^{(N-1)}\big)
    + 
    \tilde{p}^{(N-1)} 
    + 
    \lambda^{(N)} \big(x^{\text{LT},(N)} + u^{(N)}\big).
\end{align*}
Since \(x^{\text{LT},(N)} = Q^{(N)}\), we obtain
\begin{align*}
    p^{(N)} x^{\text{LT},(N)}
    =
    \left(
        \tilde{p}^{(N-1)} 
        - 
        \alpha \lambda^{(N-1)} \big(Q^{(N-1)} - Q^{(N)} + u^{(N-1)}\big)
        + 
        \lambda^{(N)} \big(Q^{(N)} + u^{(N)}\big)
    \right) Q^{(N)}.
\end{align*}
Taking conditional expectation \(\mathbb{E}_N[\cdot]\) and using 
\(\mathbb{E}_N[u^{(N)}] = 0\), we obtain
\begin{align*}
    \mathbb{E}_N\left[p^{(N)} x^{\text{LT},(N)}\right]
    =
    \left(
        \tilde{p}^{(N-1)} 
        - 
        \alpha \lambda^{(N-1)} \big(Q^{(N-1)} + u^{(N-1)}\big)
    \right) Q^{(N)}
    +
    \big(\alpha \lambda^{(N-1)} + \lambda^{(N)}\big) \big(Q^{(N)}\big)^2.
\end{align*}
Hence,
\begin{align*}
    L^{(N)}(\tilde{p}^{(N-1)}, Q^{(N-1)}, Q^{(N)})
    &=
    \left(
        \tilde{p}^{(N-1)} 
        - 
        \alpha \lambda^{(N-1)} \big(Q^{(N-1)} + u^{(N-1)}\big)
    \right) Q^{(N)} \\
    &\quad +
    \underbrace{
        \big(\alpha \lambda^{(N-1)} + \lambda^{(N)} + \phi\big)
    }_{=: \mu^{(N)}}
    \big(Q^{(N)}\big)^2.
\end{align*}

\paragraph{Inductive Step}
Assume the representation holds for periods \(N, N-1, \ldots, n+1\), i.e.,
\begin{align}
    L^{(n+1)}(\tilde{p}^{(n)}, Q^{(n)}, Q^{(n+1)})
    =
    \left(
        \tilde{p}^{(n)} 
        - 
        \alpha \lambda^{(n)} \big(Q^{(n)} + u^{(n)}\big)
    \right) Q^{(n+1)}
    +
    \mu^{(n+1)} \big(Q^{(n+1)}\big)^2.
\end{align}
The Bellman equation at time \(n\) is
\begin{align*}
    L^{(n)}(\tilde{p}^{(n-1)}, Q^{(n-1)}, Q^{(n)})
    &=
    \min_{x^{\text{LT},(n)} \in \mathbf{M}(H^{(n)})}
    \mathbb{E}_n \Big[
        p^{(n)} x^{\text{LT},(n)} 
        + 
        \phi \big(Q^{(n)}\big)^2 \\
        &\qquad\qquad\qquad
        + 
        L^{(n+1)}(\tilde{p}^{(n)}, Q^{(n)}, Q^{(n+1)})
    \Big].
\end{align*}
After substituting 
\(p^{(n)}, \tilde{p}^{(n)}, Q^{(n+1)} = Q^{(n)} - x^{\text{LT},(n)}\),
and collecting terms, we obtain a quadratic form in \(x^{\text{LT},(n)}\):
\begin{align*}
    L^{(n)} 
    =
    \min_{x^{\text{LT},(n)}}
    \Big\{
        \mu^{(n+1)} \big(x^{\text{LT},(n)}\big)^2
        +
        \big(
            \lambda^{(n)} (1+\alpha) Q^{(n)}
            - 
            2 \mu^{(n+1)} Q^{(n)}
        \big)
        x^{\text{LT},(n)} \\
        +
        \big(
            \alpha \lambda^{(n-1)} 
            + 
            \lambda^{(n)} 
            + 
            \phi
            - 
            \frac{(\lambda^{(n)})^2 (1+\alpha)^2}{4 \mu^{(n+1)}}
        \big)
        \big(Q^{(n)}\big)^2
        +
        \left(
            \tilde{p}^{(n-1)} 
            - 
            \alpha \lambda^{(n-1)} \big(Q^{(n-1)} + u^{(n-1)}\big)
        \right)
        Q^{(n)}
    \Big\}.
\end{align*}
Since \(\mu^{(n+1)} > 0\), this quadratic is strictly convex in \(x^{\text{LT},(n)}\), yielding the unique minimizer
\begin{align}
    x^{\text{LT},(n)*}
    =
    \underbrace{
        \frac{2 \mu^{(n+1)} - \lambda^{(n)} (1+\alpha)}{2 \mu^{(n+1)}}
    }_{=: \theta^{(n)}}
    Q^{(n)}.
\end{align}
Substituting this optimizer back yields
\begin{align}
     L^{(n)}(\tilde{p}^{(n-1)}, Q^{(n-1)}, Q^{(n)})
     =
     \left(
         \tilde{p}^{(n-1)} 
         - 
         \alpha \lambda^{(n-1)} \big(Q^{(n-1)} + u^{(n-1)}\big)
     \right) Q^{(n)}
     +
     \mu^{(n)} \big(Q^{(n)}\big)^2,
\end{align}
with
\begin{align}
    \mu^{(n)}
    =
    \alpha \lambda^{(n-1)}
    +
    \lambda^{(n)}
    +
    \phi
    -
    \frac{ (\lambda^{(n)})^2 (1+\alpha)^2 }{ 4 \mu^{(n+1)} }.
\end{align}
This completes the induction.
\end{proof}

\subsection{Proof of Proposition~\ref{method:claim-unanimous-VWAP}}\label{appendix:proof-unanimous-VWAP}
\begin{proof}
Substituting the pro-rata allocation rule into the VWAP definition,
\begin{align*}
    p^{(n)}
    &=
    \frac{\sum_{i = 1}^M \mathrm{Order}_i^{(n)} p_i^{(n)}}
         {\sum_{i = 1}^M \mathrm{Order}_i^{(n)}} \\
    &=
    \frac{\sum_{i = 1}^M 
        q^{(n)} 
        \frac{d_i^{(n)}}{\sum_{j=1}^M d_j^{(n)}} 
        p_i^{(n)}
    }
    {q^{(n)}} \\
    &=
    \frac{\sum_{i = 1}^M d_i^{(n)} p_i^{(n)}}
         {\sum_{i = 1}^M d_i^{(n)}}.
\end{align*}
Hence the factor \(q^{(n)}\) cancels out, showing that the execution price \(p^{(n)}\) is independent of the scale of order flow and is identical for all traders.
\\
Moreover, under the linear quoting rule for each market maker $i=1,\ldots,M$,
\[
    p_i^{(n)}
    =
    \hat{p}^{(n-1)} + \lambda_i^{(n)} q^{(n)},
\]
the VWAP can be written as
\begin{align*}
    p^{(n)}
    &=
    \frac{\sum_{i = 1}^M p_i^{(n)} d_i^{(n)}}
         {\sum_{i = 1}^M d_i^{(n)}} \\
    &=
    \frac{\sum_{i = 1}^M 
        \big(\hat{p}^{(n-1)} + \lambda_i^{(n)} q^{(n)}\big)
        d_i^{(n)}
    }
    {\sum_{i = 1}^M d_i^{(n)}} \\
    &=
    \hat{p}^{(n-1)}
    +
    q^{(n)}
    \frac{\sum_{i = 1}^M \lambda_i^{(n)} d_i^{(n)}}
         {\sum_{i = 1}^M d_i^{(n)}}.
\end{align*}
Therefore the effective impact coefficient is
\begin{equation}
    \lambda_{\mathrm{eff}}^{(n)}
    =
    \frac{\sum_{i = 1}^M \lambda_i^{(n)} d_i^{(n)}}
         {\sum_{i = 1}^M d_i^{(n)}}.
\end{equation}
Since \(d_i^{(n)} = 1/|\lambda_i^{(n)}|\), this can also be written as
\begin{equation}
    \lambda_{\mathrm{eff}}^{(n)}
    =
    \frac{\sum_{i = 1}^M \mathrm{sign}\!\left(\lambda_i^{(n)}\right)}
         {\sum_{i = 1}^M \frac{1}{|\lambda_i^{(n)}|}}.
\end{equation}
Thus, \(\lambda_{\mathrm{eff}}^{(n)}\) is a depth-weighted aggregation of individual impact parameters and is fully determined by the market makers' quoting policies at period \(n\).

\end{proof}

\subsection{Proof of Lemma~\ref{method:lemma-zero-sum-game}}
\label{appendix:proof-zero-sum-game}

\begin{proof}
Summing \eqref{eq:seq-kyle-reward-mm} over \(i\),
\begin{align*}
    \sum_{i=1}^M R_i^{\text{MM},(n)}
    &=
    \sum_{i=1}^M 
    \mathrm{Order}_i^{(n)} p_i^{(n)}
    -
    p^{(n)}
    \sum_{i=1}^M \mathrm{Order}_i^{(n)}.
\end{align*}
By definition of VWAP,
\[
    p^{(n)}
    =
    \frac{\sum_{i=1}^M \mathrm{Order}_i^{(n)} p_i^{(n)}}
         {\sum_{i=1}^M \mathrm{Order}_i^{(n)}}.
\]
Therefore,
\[
    \sum_{i=1}^M R_i^{\text{MM},(n)}
    =
    \left(
        \sum_{i=1}^M \mathrm{Order}_i^{(n)} p_i^{(n)}
    \right)
    -
    \left(
        \sum_{i=1}^M \mathrm{Order}_i^{(n)} p_i^{(n)}
    \right)
    =
    0.
\]
\end{proof}

\subsection{Characterization of Kyle model’s linear recursive equilibrium}\label{appendix:kyle-solution}

We restate the main result of the discrete-time Kyle model \citep{kyle_continuous_1985} using the notation adopted in this paper.

\begin{theorem}[Recursive Linear Equilibrium of the Kyle Model] \label{theorem: kyle}
There exists a unique linear equilibrium, which is a recursive linear equilibrium. In this equilibrium, there are constants
\(\beta^{(n)}, \lambda^{(n)}, \alpha^{(n)}, \delta^{(n)},\) and \(\Sigma^{(n)}\)
such that, for each trading period \( n \in \{1, \ldots, N\} \),
\begin{align}
    &x^{\text{IT}, (n)} = \beta^{(n)} \left( v - p^{(n - 1)} \right) \Delta t^{(n)}, \label{eq:kyle-x}\\[3pt]
    &p^{(n)} = p^{(n - 1)} + \lambda^{(n)} \left( x^{\text{IT}, (n)} + u^{(n)} \right), \label{eq:kyle-p}\\[3pt]
    &\Sigma^{(n)} := \text{Var}\!\left(v \mid q^{(1)}, \ldots, q^{(n)}\right), \label{eq:kyle-sigma}\\[3pt]
    &\mathbb{E}\!\left[ \sum_{k = n}^{N} x^{\text{IT}, (k)} (v - p^{(k)}) \,\middle|\, p^{(1)}, \ldots, p^{(n-1)}, v \right] = \alpha^{(n-1)} (v - p^{(n-1)})^2 + \delta^{(n-1)}, \label{eq:kyle-thm}
\end{align}
where \(\Delta t^{(n)} := t^{(n)} - t^{(n-1)} = \tau\) denotes the time step size.

Given \(\Sigma^{(0)}\), the constants 
\(\beta^{(n)}, \lambda^{(n)}, \alpha^{(n)}, \delta^{(n)}, \Sigma^{(n)}\)
are the unique solution to the recursive system:
\begin{align}
    \alpha^{(n-1)} &= \frac{1}{4 \lambda^{(n)} \left( 1 - \alpha^{(n)} \lambda^{(n)} \right)}, \label{eq:kyle-alpha}\\[3pt]
    \delta^{(n-1)} &= \delta^{(n)} + \alpha^{(n)} \left( \lambda^{(n)} \right)^2 \sigma_u^2 \Delta t^{(n)}, \label{eq:kyle-delta}\\[3pt]
    \beta^{(n)} \Delta t^{(n)} &= \frac{1 - 2 \alpha^{(n)} \lambda^{(n)}}{2 \lambda^{(n)} \left( 1 - \alpha^{(n)} \lambda^{(n)} \right)}, \label{eq:kyle-beta}\\[3pt]
    \lambda^{(n)} &= \frac{\beta^{(n)} \Sigma^{(n)}}{\sigma_u^2}, \label{eq:kyle-lambda}\\[3pt]
    \Sigma^{(n)} &= \left( 1 - \beta^{(n)} \lambda^{(n)} \Delta t^{(n)} \right) \Sigma^{(n-1)}, \label{eq:kyle-thm-diff-eqn}
\end{align}
subject to \(\alpha^{(N)} = \delta^{(N)} = 0\) and the second-order condition
\begin{align}
    \lambda^{(n)}(1 - \alpha^{(n)} \lambda^{(n)}) > 0.
\end{align}
\end{theorem}

\bibliographystyle{rQUF}
\bibliography{references}

@book{cartea_algorithmic_2015,
  author    = {Alvaro Cartea and Sebastian Jaimungal and José Penalva},
  title     = {Algorithmic and High-Frequency Trading},
  year      = {2015},
  publisher = {Cambridge University Press}
}

@article{alfonsi2010optimal,
  title={Optimal execution strategies in limit order books with general shape functions},
  author={Alfonsi, Aur{\'e}lien and Fruth, Antje and Schied, Alexander},
  journal={Quantitative finance},
  volume={10},
  number={2},
  pages={143--157},
  year={2010},
  publisher={Taylor \& Francis}
}

@article{su2022decentralized,
  title={Decentralized policy optimization},
  author={Su, Kefan and Lu, Zongqing},
  journal={arXiv preprint arXiv:2211.03032},
  year={2022}
}

@inproceedings{jin2023stationary,
  title={On stationary point convergence of PPO-Clip},
  author={Jin, Ruinan and Li, Shuai and Wang, Baoxiang},
  booktitle={The Twelfth International Conference on Learning Representations},
  year={2023}
}

@inproceedings{huang2024ppo,
  title={Ppo-clip attains global optimality: Towards deeper understandings of clipping},
  author={Huang, Nai-Chieh and Hsieh, Ping-Chun and Ho, Kuo-Hao and Wu, I-Chen},
  booktitle={Proceedings of the AAAI Conference on Artificial Intelligence},
  volume={38},
  number={11},
  pages={12600--12607},
  year={2024}
}

@article{kyle_continuous_1985,
  author    = {Albert S. Kyle},
  title     = {Continuous auctions and insider trading},
  journal   = {Econometrica},
  volume    = {53},
  number    = {6},
  pages     = {1315--1335},
  year      = {1985}
}

@article{glosten_bid_1985,
  author    = {Lawrence R. Glosten and Paul R. Milgrom},
  title     = {Bid, ask and transaction prices in a specialist market with heterogeneously informed traders},
  journal   = {Journal of Financial Economics},
  volume    = {14},
  number    = {1},
  pages     = {71--100},
  year      = {1985},
  doi       = {10.1016/0304-405X(85)90044-3}
}

@article{Ho_1981,
  author    = {Thomas Ho and Hans R. Stoll},
  title     = {Optimal dealer pricing under transactions and return uncertainty},
  journal   = {Journal of Financial Economics},
  volume    = {9},
  number    = {1},
  pages     = {47--73},
  year      = {1981},
  doi       = {10.1016/0304-405X(81)90020-9}
}

@article{schulman_proximal_2017,
  author        = {John Schulman and Filip Wolski and Prafulla Dhariwal and Alec Radford and Oleg Klimov},
  title         = {Proximal policy optimization algorithms},
  journal   = {arXiv preprint arXiv:1707.06347},
  year          = {2017},
  eprint        = {1707.06347},
  archivePrefix = {arXiv},
  primaryClass  = {cs.LG},
  url           = {https://arxiv.org/abs/1707.06347}
}

@article{lowe_multi-agent_2017,
  author        = {Ryan Lowe and Yi Wu and Aviv Tamar and Jean Harb and Pieter Abbeel and Igor Mordatch},
  title         = {Multi-agent actor-critic for mixed cooperative-competitive environments},
  journal   = {arXiv preprint arXiv:1706.02275},
  year          = {2017},
  eprint        = {1706.02275},
  archivePrefix = {arXiv},
  primaryClass  = {cs.LG},
  url           = {https://arxiv.org/abs/1706.02275}
}

@book{byrd_abides_2019,
  author    = {David Byrd and Maria Hybinette and Tucker Balch},
  title     = {ABIDES: Towards high-fidelity multi-agent market simulation},
  booktitle = {Proceedings of the 2020 ACM SIGSIM Conference on Principles of Advanced Discrete Simulation},
  year      = {2020},
  pages     = {11--22},
  doi       = {10.1145/3384441.3395986},
  publisher = {Association for Computing Machinery}
}

@inproceedings{amrouni_abides-gym_2021,
  author        = {Selim Amrouni and Aymeric Moulin and Jared Vann and Svitlana Vyetrenko and Tucker Balch and Manuela Veloso},
  title         = {ABIDES-Gym: Gym environments for multi-agent discrete event simulation and application to financial markets},
  booktitle     = {Second ACM International Conference on AI in Finance},
  series        = {ICAIF'21},
  publisher     = {ACM},
  year          = {2021},
  pages         = {1--9},
  doi           = {10.1145/3490354.3494433}
}

@article{almgren_optimal_2001,
  author    = {Robert Almgren and Neil Chriss},
  title     = {Optimal execution of portfolio transactions},
  journal   = {The Journal of Risk},
  volume    = {3},
  number    = {2},
  pages     = {5--39},
  year      = {2001},
  doi       = {10.21314/JOR.2001.041}
}

@article{BERTSIMAS1998,
  author    = {Dimitris Bertsimas and Andrew W. Lo},
  title     = {Optimal control of execution costs},
  journal   = {Journal of Financial Markets},
  volume    = {1},
  number    = {1},
  pages     = {1--50},
  year      = {1998},
  doi       = {10.1016/S1386-4181(97)00012-8}
}

@inproceedings{yu_surprising_2022,
  author    = {Chao Yu and Akash Velu and Eugene Vinitsky and Jiaxuan Gao and Yu Wang and Alexandre Bayen and Yi Wu},
  title     = {The surprising effectiveness of PPO in cooperative, multi-agent games},
  booktitle = {Advances in Neural Information Processing Systems},
  volume    = {35},
  year      = {2022}
}

@article{kuba_trust_2022,
  author        = {Jakub Grudzien Kuba and Ruiqing Chen and Muning Wen and Ying Wen and Fanglei Sun and Jun Wang and Yaodong Yang},
  title         = {Trust region policy optimisation in multi-agent reinforcement learning},
  journal   = {arXiv preprint arXiv:2109.11251},
  year          = {2022},
  eprint        = {2109.11251},
  archivePrefix = {arXiv},
  primaryClass  = {cs.AI},
  url           = {https://arxiv.org/abs/2109.11251}
}

@inproceedings{Karpe_2020,
  author        = {Michaël Karpe and Jin Fang and Zhongyao Ma and Chen Wang},
  title         = {Multi-agent reinforcement learning in a realistic limit order book market simulation},
  booktitle     = {First ACM International Conference on AI in Finance},
  publisher     = {ACM},
  year          = {2020},
  articleno     = {30},
  numpages      = {7},
  doi           = {10.1145/3383455.3422570},
  url           = {https://doi.org/10.1145/3383455.3422570}
}

@inproceedings{nevmyvaka2006reinforcement,
  author    = {Yuriy Nevmyvaka and Yi Feng and Michael Kearns},
  title     = {Reinforcement learning for optimized trade execution},
  booktitle = {23rd International Conference on Machine Learning},
  pages     = {673--680},
  year      = {2006}
}

@inproceedings{hendricks2014reinforcement,
  author    = {Dieter Hendricks and Diane Wilcox},
  title     = {A reinforcement learning extension to the Almgren--Chriss framework for optimal trade execution},
  booktitle = {IEEE Conference on Computational Intelligence for Financial Engineering and Economics (CIFEr)},
  pages     = {457--464},
  year      = {2014},
  publisher = {IEEE},
  doi       = {10.1109/CIFER.2014.6924109}
}

@article{friedrich2020deep,
  author = {Paul Friedrich and Josef Teichmann},
  title  = {Deep investing in Kyle's single period model},
  journal = {arXiv preprint arXiv:2006.13889},
  year    = {2020},
  eprint  = {2006.13889},
  archivePrefix = {arXiv},
  primaryClass  = {q-fin.TR},
  url     = {https://arxiv.org/abs/2006.13889}
}

@article{ganesh2019reinforcement,
  author    = {Apoorv Ganesh and Gabriele D’Adamo and Naveed Firoozye and Stylianos Skoulakis},
  title     = {Reinforcement learning for market making in a multi-agent dealer market},
  journal   = {arXiv preprint arXiv:1911.05892},
  year      = {2019},
  eprint    = {1911.05892},
  archivePrefix = {arXiv},
  primaryClass  = {q-fin.TR},
  url       = {https://arxiv.org/abs/1911.05892}
}

@article{hafsi2024optimalexecutionreinforcementlearning,
  author        = {Yadh Hafsi and Edoardo Vittori},
  title         = {Optimal execution with reinforcement learning},
  year          = {2024},
  journal   = {arXiv preprint arXiv:2411.06389},
  eprint        = {2411.06389},
  archivePrefix = {arXiv},
  primaryClass  = {q-fin.TR},
  url           = {https://arxiv.org/abs/2411.06389}
}

@article{Nagy_2023,
  author    = {Peer Nagy and Jan-Peter Calliess and Stefan Zohren},
  title     = {Asynchronous deep double dueling Q-learning for trading-signal execution in limit order book markets},
  journal   = {Frontiers in Artificial Intelligence},
  volume    = {6},
  year      = {2023},
  doi       = {10.3389/frai.2023.1151003},
  url       = {https://www.frontiersin.org/articles/10.3389/frai.2023.1151003}
}

@article{Huberman2005Liquidity,
  author    = {Gur Huberman and Werner Stanzl},
  title     = {Optimal liquidity trading},
  journal   = {Review of Finance},
  volume    = {9},
  number    = {2},
  pages     = {165--200},
  year      = {2005},
  doi       = {10.1007/s10679-005-7591-5}
}

@article{Huberman2004Manipulation,
  author    = {Gur Huberman and Werner Stanzl},
  title     = {Price manipulation and quasi-arbitrage},
  journal   = {Econometrica},
  volume    = {72},
  number    = {4},
  pages     = {1247--1275},
  year      = {2004}
}

@article{Ning2018DDQNForOptExecution,
  author      = {Brian Ning and Franco Ho Ting Lin and Sebastian Jaimungal},
  title   = {Double deep Q-learning for optimal execution},
  journal = {Applied Mathematical Finance},
  volume  = {27},
  number  = {3},
  pages   = {173--199},
  year    = {2021},
  doi     = {10.1080/1350486X.2020.1783194}
}

@inproceedings{Lin2020,
  author    = {Siyu Lin and Peter A. Beling},
  title     = {An end-to-end optimal trade execution framework based on proximal policy optimization},
  booktitle = {Twenty-Ninth International Joint Conference on Artificial Intelligence (IJCAI-20)},
  pages     = {4548--4554},
  year      = {2020},
  doi       = {10.24963/ijcai.2020/627}
}

@book{Agent-Based-Market-Simulator,
  author    = {Siyu Lin and Peter A. Beling},
  title     = {An agent-based market simulator for back-testing deep reinforcement learning based trade execution strategies},
  booktitle = {Neural Information Processing},
  publisher = {Springer International Publishing},
  year      = {2021},
  pages     = {644--653}
}

@article{Jaisson2022,
  author    = {Thibault Jaisson},
  title     = {Deep differentiable reinforcement learning and optimal trading},
  journal   = {Quantitative Finance},
  volume    = {22},
  number    = {8},
  pages     = {1429--1443},
  year      = {2022},
  doi       = {10.1080/14697688.2022.2062431}
}

@article{LiEtAl2024,
  author    = {Fengpei Li and Vitalii Ihnatiuk and Yu Chen and Jiahe Lin and Ryan J. Kinnear and Anderson Schneider and Yuriy Nevmyvaka and Henry Lam},
  title     = {Do price trajectory data increase the efficiency of market impact estimation?},
  journal   = {Quantitative Finance},
  volume    = {24},
  number    = {5},
  pages     = {545--568},
  year      = {2024},
  doi       = {10.1080/14697688.2024.2351457}
}

@article{DupretHainaut2025,
  author    = {Jean-Loup Dupret and Donatien Hainaut},
  title     = {Optimal liquidation under indirect price impact with propagator},
  journal   = {Quantitative Finance},
  volume    = {25},
  number    = {3},
  pages     = {359--381},
  year      = {2025},
  doi       = {10.1080/14697688.2025.2463368}
}

@article{MoallemiWang2022,
  author    = {Ciamac C. Moallemi and Muye Wang},
  title     = {A reinforcement learning approach to optimal execution},
  journal   = {Quantitative Finance},
  volume    = {22},
  number    = {6},
  pages     = {1051--1069},
  year      = {2022},
  doi       = {10.1080/14697688.2022.2039403}
}

@article{Gatheral2010,
  author    = {Jim Gatheral},
  title     = {No-dynamic-arbitrage and market impact},
  journal   = {Quantitative Finance},
  volume    = {10},
  number    = {7},
  pages     = {749--759},
  year      = {2010},
  doi       = {10.1080/14697680903373692}
}

@article{ObizhaevaWang2013,
  author    = {Anna A. Obizhaeva and Jiang Wang},
  title     = {Optimal trading strategy and supply/demand dynamics},
  journal   = {Journal of Financial Markets},
  volume    = {16},
  number    = {1},
  pages     = {1--32},
  year      = {2013},
  doi       = {10.1016/j.finmar.2012.09.001}
}

@article{Curato2017,
  author    = {Gianbiagio Curato and Jim Gatheral and Fabrizio Lillo},
  title     = {Optimal execution with non-linear transient market impact},
  journal   = {Quantitative Finance},
  volume    = {17},
  number    = {1},
  pages     = {41--54},
  year      = {2017},
  doi       = {10.1080/14697688.2016.1181274}
}

@article{cheridito2025reinforcementlearningtradeexecution,
  author        = {Patrick Cheridito and Moritz Weiss},
  title         = {Reinforcement learning for trade execution with market impact},
  journal       = {arXiv preprint arXiv:2507.06345},
  year          = {2025},
  eprint        = {2507.06345},
  archivePrefix = {arXiv},
  primaryClass  = {q-fin.TR},
  url           = {https://arxiv.org/abs/2507.06345}
}

@article{espana2025reinforcementlearningqueuereactivemodels,
      title={Reinforcement learning in queue-reactive models: application to optimal execution}, 
      journal={arXiv preprint arXiv:2511.15262},
      author={Tomas Espana and Yadh Hafsi and Fabrizio Lillo and Edoardo Vittori},
      year={2025},
      eprint={2511.15262},
      archivePrefix={arXiv},
      primaryClass={q-fin.TR},
      url={https://arxiv.org/abs/2511.15262}, 
}

@article{CheriditoSepin2014,
  author  = {Cheridito, Patrick and Sepin, Tardu},
  title   = {Optimal trade execution under stochastic volatility and liquidity},
  journal = {Applied Mathematical Finance},
  volume  = {21},
  number  = {4},
  pages   = {342--362},
  year    = {2014},
  doi     = {10.1080/1350486X.2014.881005}
}

\end{document}